\documentclass[fleqn,usenatbib]{mnras}

\usepackage[T1]{fontenc}

\DeclareRobustCommand{\VAN}[3]{#2}
\let\VANthebibliography\thebibliography
\def\thebibliography{\DeclareRobustCommand{\VAN}[3]{##3}\VANthebibliography}

\usepackage{graphicx}	
\usepackage{amsmath}	
\usepackage{amssymb}	
\usepackage{lscape}
\usepackage{ulem}
\usepackage{tabularx}
\usepackage[dvipsnames]{xcolor}
\usepackage{soul}
\usepackage{caption}
\usepackage{subcaption}
\usepackage{tikz}

\graphicspath{{images/}}

\title[HyMoR galaxies from the FIRST survey]{Search for hybrid morphology radio galaxies from the FIRST survey at 1400 MHz}

\author[S. Kumari and S. Pal]{
Shobha Kumari$^{1}$\orcidA{}
Sabyasachi Pal$^{1}$\orcidB\thanks{E-mail: sabya.pal@gmail.com (SP)}\\
$^{1}$Midnapore City College, Kuturia, Bhadutala, Paschim Medinipur, West Bengal, 721129, India \\
}

\newcommand{\orcidicon}{\includegraphics[width=0.32cm]{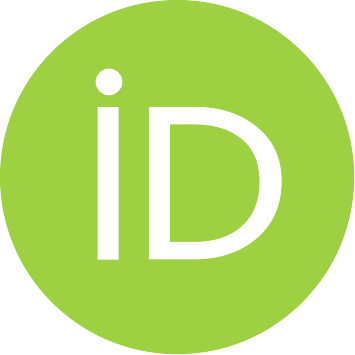}}

\foreach \x in {A, ..., Z}{%
\expandafter\xdef\csname orcid\x\endcsname{\noexpand\href{https://orcid.org/\csname orcidauthor\x\endcsname}{\noexpand\orcidicon}}
}

\date{Accepted 2022 April 22. Received 2022 April 22; in original form 2021 April 30}

\pubyear{2022}

\usepackage{newtxtext,newtxmath}
\begin{document}
\label{firstpage}
\pagerange{\pageref{firstpage}--\pageref{lastpage}}
\maketitle
\begin{abstract}
Hybrid Morphology Radio Sources (HyMoRS) are a very rare subclass of radio galaxies with apparent mixed FR morphology, i.e. these galaxies seem to have an FR-I structure on one side of the core and an FR-II structure on the other side of the core. We systematically searched for HyMoRS using Very Large Array (VLA) Faint Images of the Radio Sky at Twenty-cm (FIRST) survey with 1400 MHz frequency and identified 33 candidate HyMoRS. Our finding significantly increased the known sample size of HyMoRS. HyMoRS may play an essential role in understanding the interaction of jets with the interstellar medium and the much-debated topic of the FR dichotomy. We identified optical counterparts for 29 sources in our catalogue. In our sample of sources, one source (J1106+1355) had quasar-like behaviour. Four sources were BRCLG (Brightest Cluster Galaxies) and six were LRG (Luminous Red Galaxies). We have estimated the spectral index and radio luminosity of HyMoRS in our catalogue, when possible. We found that J1136--0328 was the most luminous source in our sample ($\log L = 27.01$ W Hz$^{-1}$sr$^{-1}$). It was also the farthest HyMoRS (with a redshift $z$ = 0.82) in our sample. With the help of a large sample size of discovered sources, various statistical properties of detected galaxies were studied.
 \end{abstract}



\section{INTRODUCTION}
\label{sec:intro}
Radio galaxies are known to have a high mass black hole ($M\sim10^{9-12}M_\odot$) in their core and two oppositely directed jets erupting from the core. Based on the FR ratio (the ratio of the distance
between the regions of strongest brightness on opposite sides of the central galaxy to the farther extent of the extremities of the source), \citet{Fa74} categorized radio galaxies as FR-I and FR-
II. FR-I sources are recognised by prominent twin jet architecture, diffuse plumes, and the strong surface brightness in each jet located towards the centre with an FR ratio of $<$0.5. These jets, which can
travel up to tens of kpc range, were originally thought to represent the major particle acceleration \citep{La99}. The jets of FR-II sources are assumed to be highly relativistic and travel long distances
(up to the order of Mpc) and are often one sided. These jets cease in a shock, which is usually placed at the outer edge of the structure. These sources are also distinguished by the radio luminosity shown by \citet{Ow94}. It was found that most of the FR-I radio galaxies have lower critical radio luminosity ($L_{\textrm{178 MHz}}<2\times10^{25}$ W Hz$^{-1}$sr$^{-1}$) than that of FR-II radio galaxies ($L_{\textrm{178 MHz}}>2\times10^{25}$ W Hz$^{-1}$sr$^{-1}$) \citep{Fa74, Ow94}.

A rare type of radio source was identified as Hybrid Morphology Radio Sources (HyMoRS) by \citet{Go00}, where the radio morphology of these sources exhibited both types of structures (FR-I on one side and FR-II on the other side of the core). The FR dichotomy has been a widely discussed topic for more than 40 yr \citep{Fa74, Ch09} and the study of hybrid radio sources might be of fundamental importance
to understanding the origin of the FR dichotomy.

These types of sources will give us an idea of how the morphology of radio galaxies is linked to the nature of the central engine, its environment, and the composition of jets \citep{Go00, Ga17}. Though the origin of HyMoRS is not completely understood, it is believed that HyMoRS might be the result of the strongly asymmetric interaction of jets with the external medium [interstellar medium (ISM) as well as intergalactic medium (IGM)] \citep{Go00, Ka17, Ha20}. It is also considered that restarted activity in SMBH, non-uniform environment, density ratio between the external dense medium, propagation of jets (due to parameters like pressure, accretion rate, and mass of SMBH), as well as doppler boosting, can have a significant impact on the observed morphology of HyMoRS \citep{Ce13, Ga17, Ka17, Ha20}. HyMoRS morphology may also be due to some projection effects \citep{Ka17, Ha20}. We need to remember that we are observing a two dimensional (2D) image of a 3D object. The effects of orientation can have a significant impact on the observed morphology. 

The first galaxy with a HyMoR like structure was reported by \citet{Sa96}, without identifying it as HyMoRS. \citet{Go00} first mentioned the HyMoR class of radio galaxy and reported six candidate HyMoRS using the Giant Metrewave Radio Telescope. Using 1700 sources from the Faint Images of the Radio Sky at Twenty-cm (FIRST) survey \citep{Be95, Wh97}, \citet{Ga06} found three certain and two possible HyMoRS. These sources are very rare ($<$1 per cent of radio galaxies belong to this category \citep{Ga06, Ka17}). So far, only a small number of HyMoRS are known, and only a small fraction of them have been studied in detail. Using VLBI observations \citet{Ce13} examined the five known HyMoRS with a 10 kpc jet length, which suggested that the origin of HyMoRS is not due to the orientation of jets as was discussed earlier \citep{Ma06}. 
This paper aims to increase the known sample size of HyMoRS by systematically inspecting the sky covered by the Very Large Array (VLA) Faint Images of the Radio Sky at Twenty-cm (FIRST) survey \citep{Be95, Wh97} and thus help to understand the nature of these sources in detail. 

This paper is organized as follows: Summary of the VLA FIRST survey, the definition of newly detected radio sources, source identification strategy, and optical counterpart identification of sources are described in Section \ref{sec:source-identification}. In Section \ref{sec:result}, we elaborated different results. In Section \ref{sec:disc}, we discussed our findings, and in Section \ref{sec:conclusion}, we summarised the conclusions.
We used the following $\Lambda$CDM cosmology parameters for the entire discussion in this paper using results from the final full-mission Planck measurements of the CMB anisotropies: $H_0$ = 67.4 km s$^{-1}$ Mpc$^{-1}$, $\Omega_{vac}$ = 0.685 and $\Omega_m$ = 0.315 \citep{Ag20}.

\section{IDENTIFYING OF THE HYBRID MORPHOLOGY RADIO SOURCES}
\label{sec:source-identification}
\subsection{The VLA FIRST survey}
\label{subsec:First}
The FIRST survey covered a radio sky of 10 575 square degrees near the North and South Galactic caps at 1400 MHz (21 cm). This survey had a typical RMS of 0.15 mJy and an angular resolution of 5 arcsec \citep{Be95, Wh97}. The FIRST survey
covered approximately 25 per cent of the total sky, out of which, approximately 80 per cent was in the North Galactic cap (8444 deg$^2$), and 20 per cent was in the South Galactic cap (2131 deg$^2$) \citep{Be95, Wh97}. The FIRST survey offered better resolution and sensitivity than the previous National Radio Astronomy Observatory (NRAO) VLA Sky Survey (NVSS) at 1.4 GHz, which used the VLA-D configuration and covered 82 per cent of the celestial sphere with an angular resolution of 45 arcsec and an RMS of $\sim$0.45 mJy \citep{Co98}. The FIRST survey used the NRAO VLA in its B-configuration (VLA-B). 

With the high sensitivity and resolution of the FIRST survey, the study of the morphology of faint radio galaxies became possible in detail. In the last two decades, the FIRST data base was used to search for radio galaxies with different distinct morphologies, such as compact steep spectrum sources, core-dominated triple sources \citep{Ku02, Ma06}, giant radio sources \citep{Ku18}, and head-tailed sources \citep{Mis19, Sa22}. Earlier, around seven hundred winged radio galaxy candidates were discovered with the help of the FIRST database \citep{Ch07, Ya19, Be20}.

\subsection{Definition of hybrid morphology radio sources}
\label{subsubsec:definition}
We classified HyMoRS based on the morphology of each source. We noted the location of the brightest peak in individual lobes of radio galaxies and recorded whether it was edge-oriented (FR-II) or centre-oriented (FR-I). We measured the FR index ($R$) based on the peaks and lobe distance of both sides of the galaxy from the core using the equation \citep{Ka17}.
\begin{equation}	
     R=2h/d+0.5
    \label{equ:FR}
\end{equation}
where $h$ is the angular distance of peaks from the core and $d$ is the angular size of the individual lobe. We catalogued the sources as HyMoRS if they have $R>1.5$ (FR-II) on one side and $0.5<R<1.5$ (FR-I) on the other side. 
 
We noted that the size of the lobe depends on the sensitivity of the map and the last contour of the image. For uniformity, we used 0.45 mJy ($\sim$3$\sigma$) as the lower contour for all 33 sources for measurement of lobe length for the HyMoRS classification purposes.

\begin{figure}	
\centerline{
\includegraphics[height=10.5cm,width=7.5cm,angle=-90]{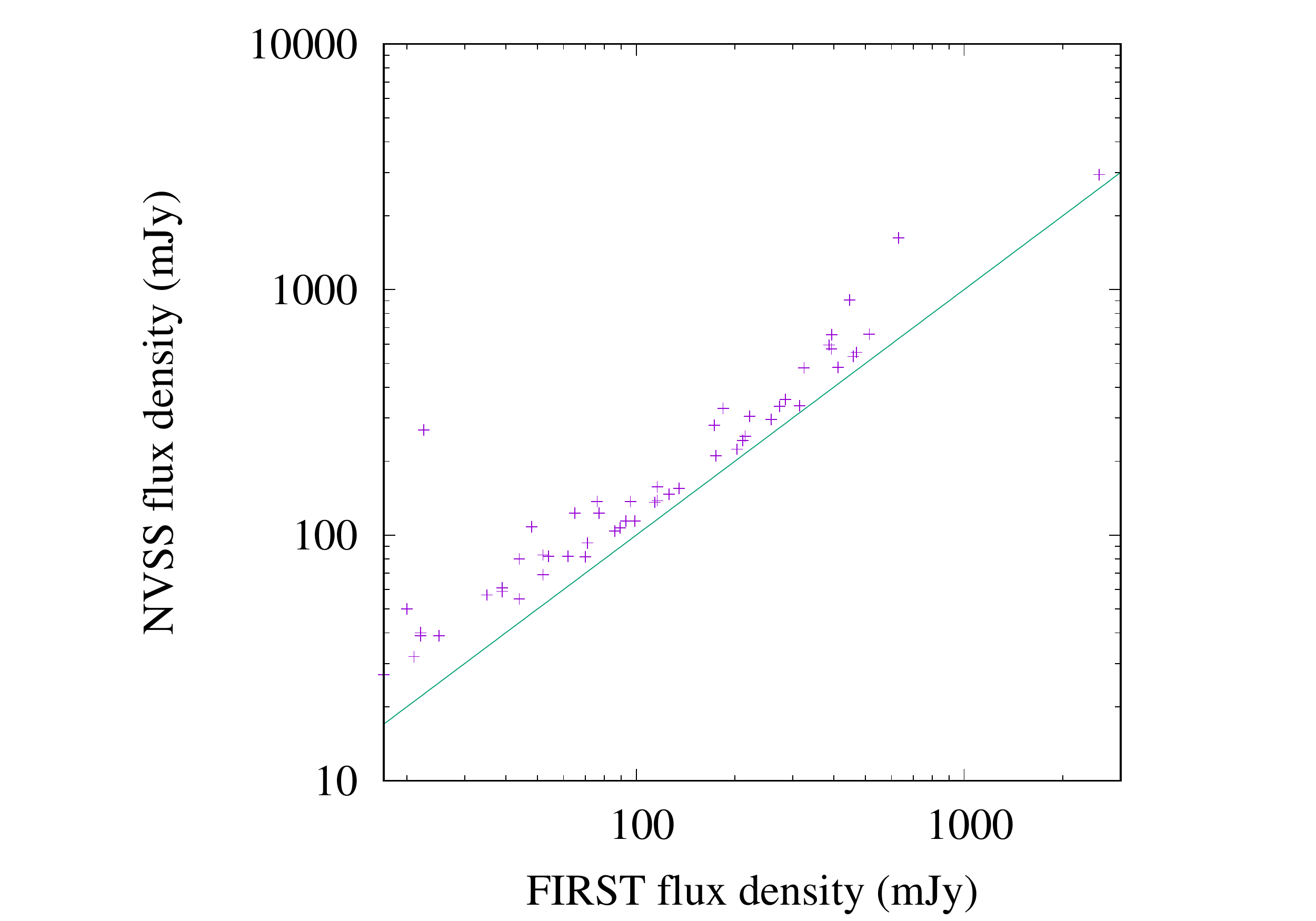}
}
\caption{Flux densities measured from the FIRST and NVSS were shown for all HyMoRS presented in this paper. Errors in flux density for FIRST and NVSS were also shown. Here, the green straight line represented the 1:1 ratio between flux densities from the FIRST and NVSS surveys.}
\label{fig:first-nvss}
\end{figure}

\subsection{Search strategy}
\label{subsec:search-strategy}
The FIRST catalogue contains a total of 946 432 radio sources. We filtered all sources in the catalogue having an angular size of $>15\arcsec$ (i.e. at least thrice the convolution beam size). Our filtering gave an output of 20 045 sources. We visually examined the morphology of all radio sources from the field of 20 045 sources and identified sources with possible FR-I/FR-II/hybrid morphology. To check whether they have a hybrid morphology structure, we measured the FR index ($R$, as defined in subsection \ref{subsubsec:definition}) of both lobes for each of the possible FR-I/FR-II/hybrid radio galaxies and checked whether they have ${\textit R}>1.5$ (FR-II) on one side and 0.5$<{\textit R}<1.5$ (FR-I) on the other side.

Through our study, we successfully discovered thirty-three HyMoRS. We removed two sources (J1224+0203 and J1034+2518) from our catalogue as they were listed earlier in \citet{Ka17}. It should be remembered that extended sources with low surface brightness will be missed by the FIRST sample completely or partially, which may restrict their identification as HyMoRS. Multicomponent large radio galaxies will not be detected by the FIRST survey due to the lack of short-spacing of uv data, for which some of the multicomponent radio galaxies may just look like point sources in the FIRST. This may lead to the missing of some of the large HyMoRS in the FIRST survey. 

We checked the NVSS images of all selected HyMoRS to look for any extended emissions. We found that for none of the sources in our list of HyMoRS, individual lobes were resolved in corresponding NVSS images.

\subsection{The optical counterparts and properties}
\label{subsubsec:optical-counterpart}
For each of the newly discovered HyMoRS, we looked for the optical counterparts using Panoramic Survey Telescope and Rapid Response System (Pan-STARRS)\footnote{https://ps1images.stsci.edu/cgi-bin/ps1cutouts} \citep{Ch16}, the Sloan Digital Sky Survey (SDSS-DR16) data catalogue\footnote{https://www.sdss.org/dr16/} \citep{Ahu20} and NED\footnote{https://ned.ipac.caltech.edu}. The identification of the optical counterpart was based on the position of the optical source relative to the radio galaxy morphology. The Pan-STARRS (red) optical filter images were overlaid with the radio images from the VLA FIRST survey. The positions of the optical counterparts of HyMoRS were used as the positions of these sources and were presented in the 3rd and 4th columns of Table \ref{tab:HyMoRS}. Optical/IR counterparts were found for 29 sources (16 were from the SDSS-DR16 \citep{Ahu20} catalogue, 10 from Pan-STARRS, and 3 were from NED) out of a total of 33 sources. The location of the core of the radio galaxy or the intersection of both radio lobes was used as the position of the galaxy when no clear optical or IR counterparts were available.
 
\section{RESULTS}
\label{sec:result}
\subsection{Newly detected hybrid radio galaxies}
We reported the discovery of 33 HyMoRS from the VLA FIRST survey, which is summarized in Table \ref{tab:HyMoRS}. We identified the optical counterparts of sources, when available and in the 5th column of Table \ref{tab:HyMoRS}, we mentioned the name of the catalogue/telescope from which the optical counterpart was identified. For flux density measurements in this paper, we used NVSS \citep{Co98} flux densities at 1400 MHz instead of flux density measurements from FIRST and tabulated them in column 6.
 
The sensitivity to low surface brightness is poor for the FIRST survey as the survey is prone to flux density loss due to its lack of antennas at short spacing (short baselines). NVSS was better suited to detecting the most extended radio structure with a better baseline distribution in short spacing. 

In Fig. \ref{fig:first-nvss}, the FIRST and NVSS flux-density distribution of all HyMoRS was shown, which confirmed that the flux densities from the NVSS images were significantly higher than the corresponding measurements from the FIRST images. The mean and median flux-density measurements from the NVSS catalogue were 279 and 136.5 mJy, and the mean and median flux densities from the measurements of the FIRST catalogue were 215.6 and 115 mJy. For one source, J0756+3901, there was a background source in the FIRST image inside the beam size of the NVSS image. For this source, we subtracted the flux density of the background point source measured from the FIRST survey from the corresponding NVSS measurement.

\begin{figure*}
\vbox{
\centerline{
\includegraphics[width=5.5cm,height=5.5cm,origin=c]{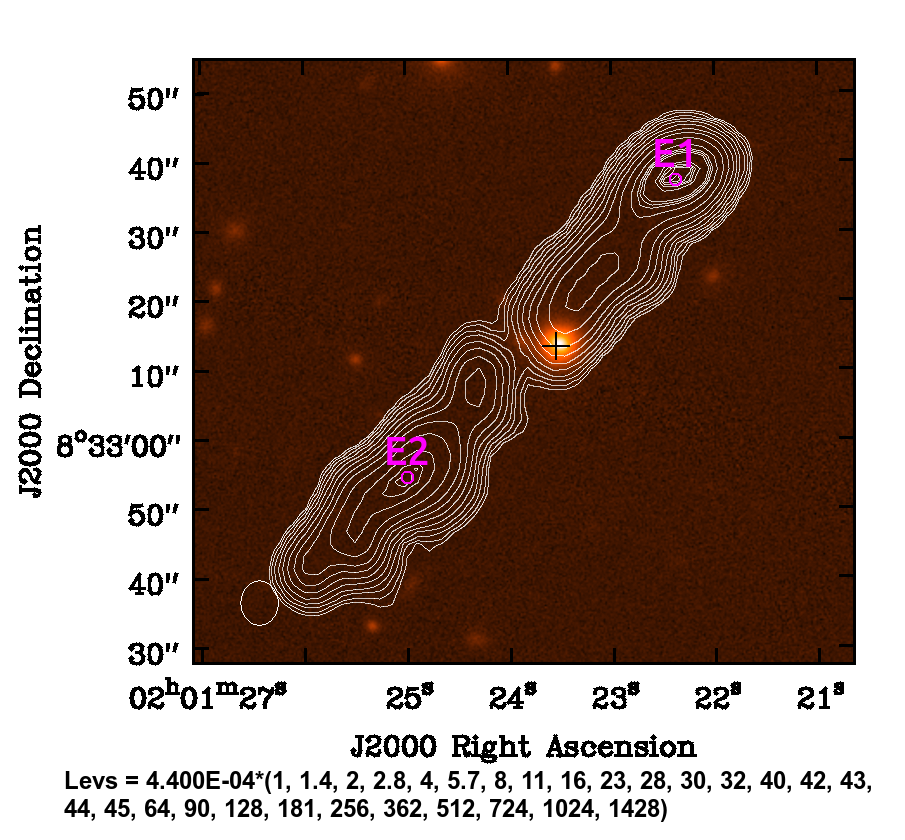}
\includegraphics[width=5.5cm,height=5.5cm,origin=c]{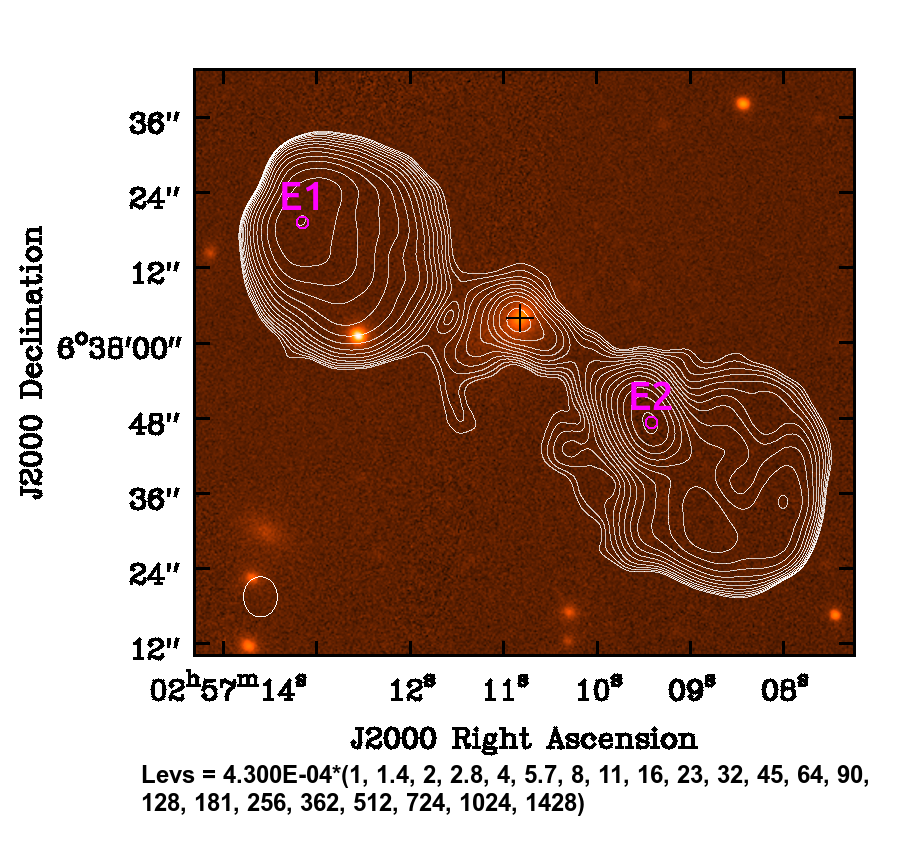}
\includegraphics[width=5.5cm,height=5.5cm,origin=c]{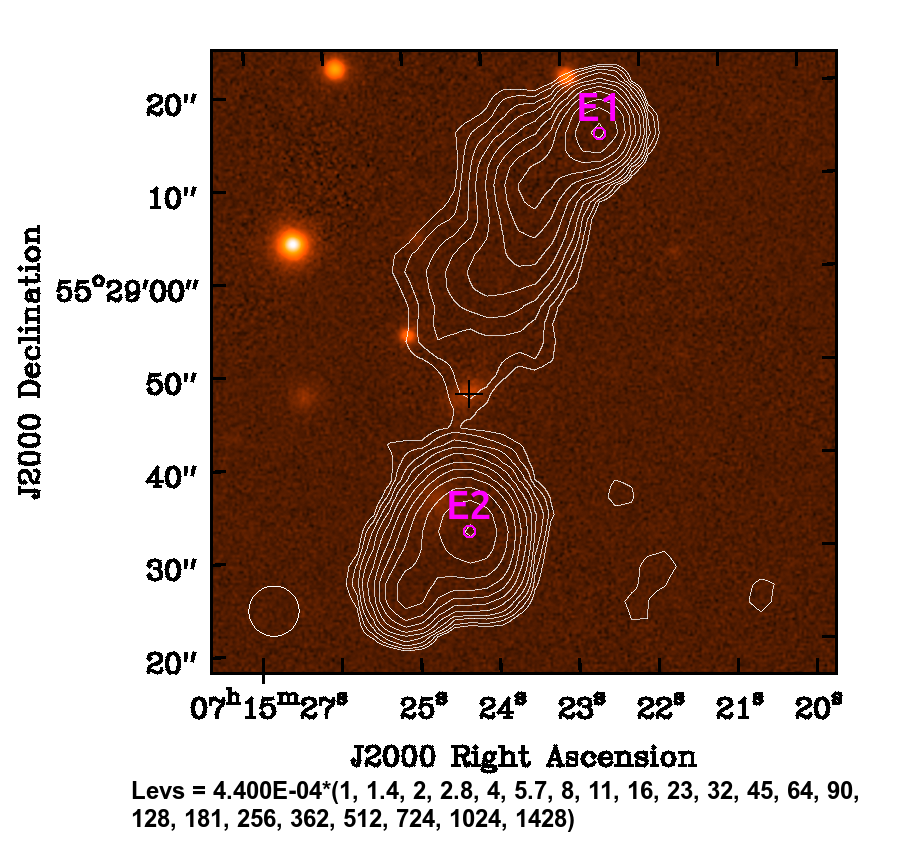}
}
}
\vbox{
\centerline{
\includegraphics[width=5.5cm,height=5.5cm,origin=c]{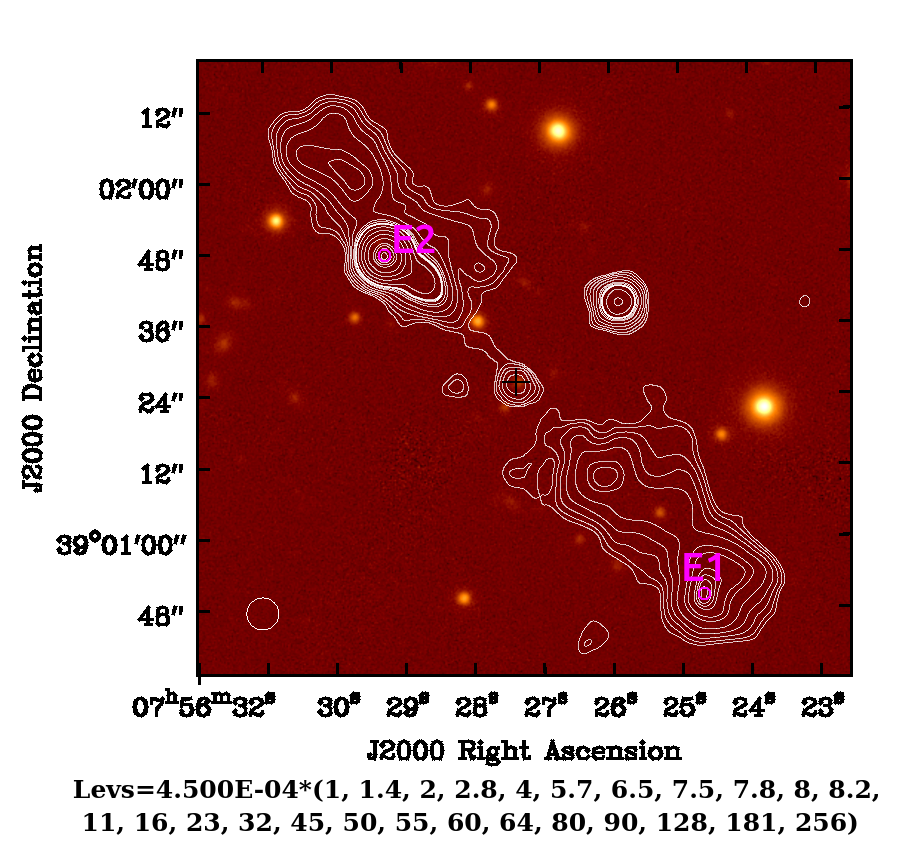}
\includegraphics[width=5.5cm,height=5.5cm,origin=c]{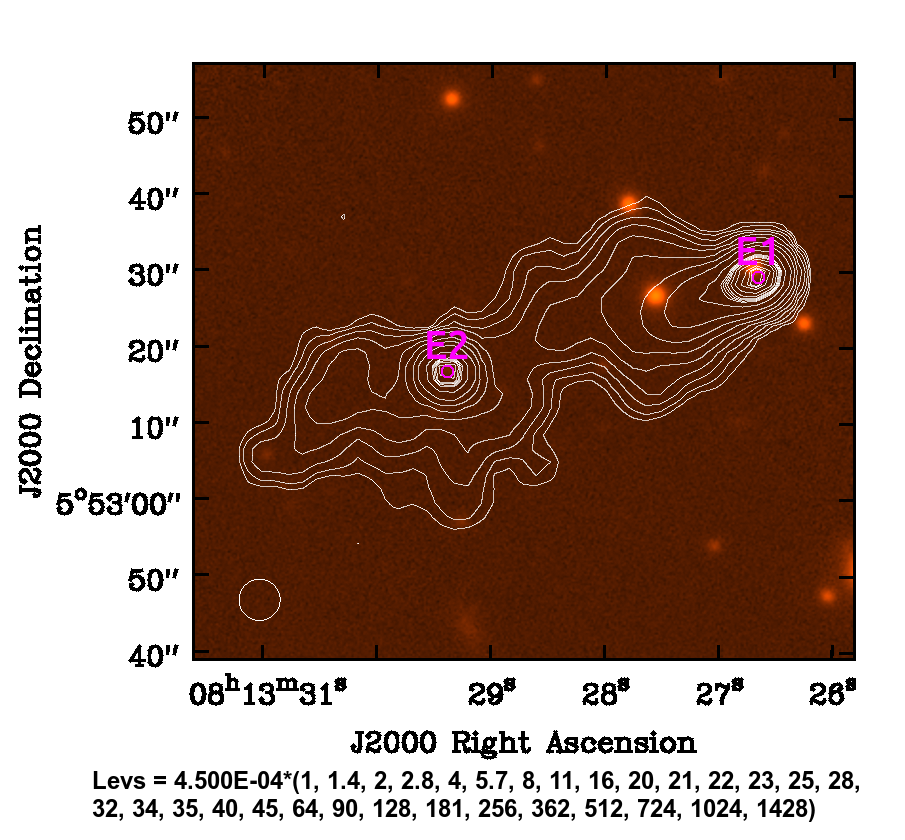}
 \includegraphics[width=5.5cm,height=5.5cm,origin=c]{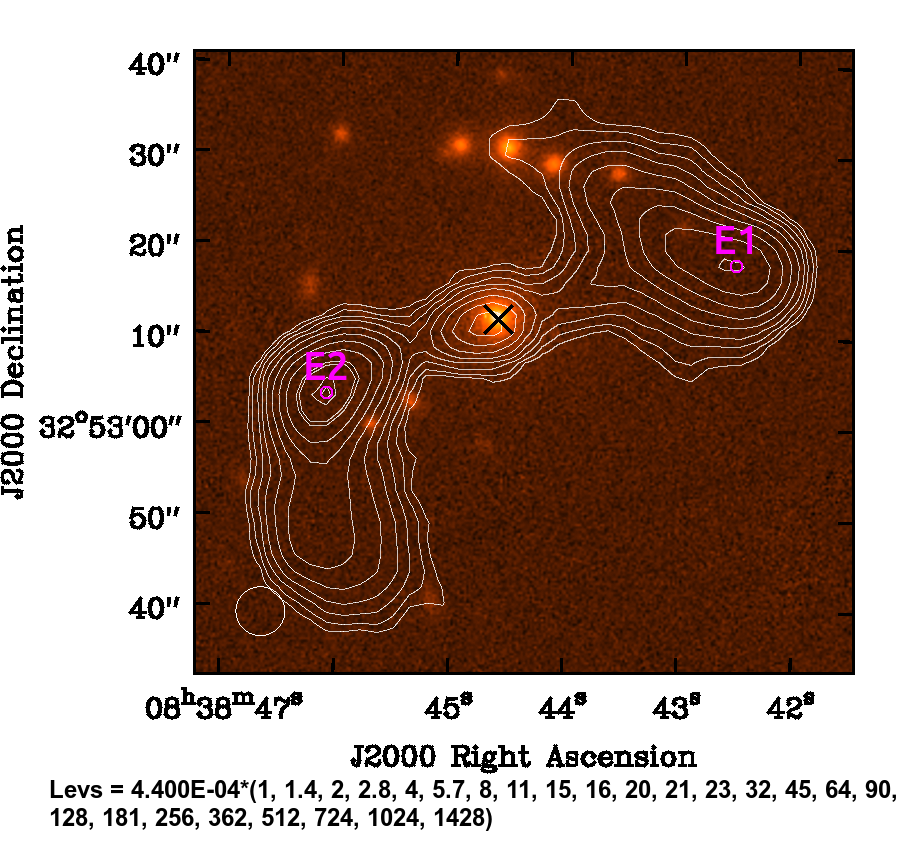}
}
}

\vbox{
\centerline{
\includegraphics[width=5.5cm,height=5.5cm,origin=c]{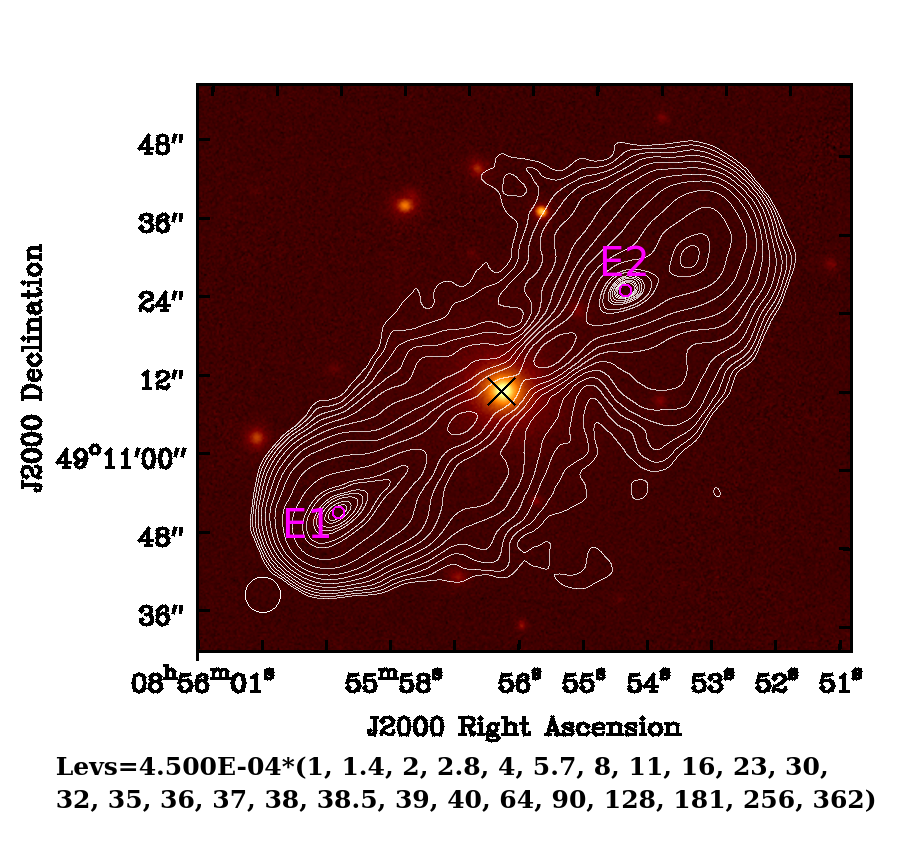}
 \includegraphics[width=5.5cm,height=5.5cm,origin=c]{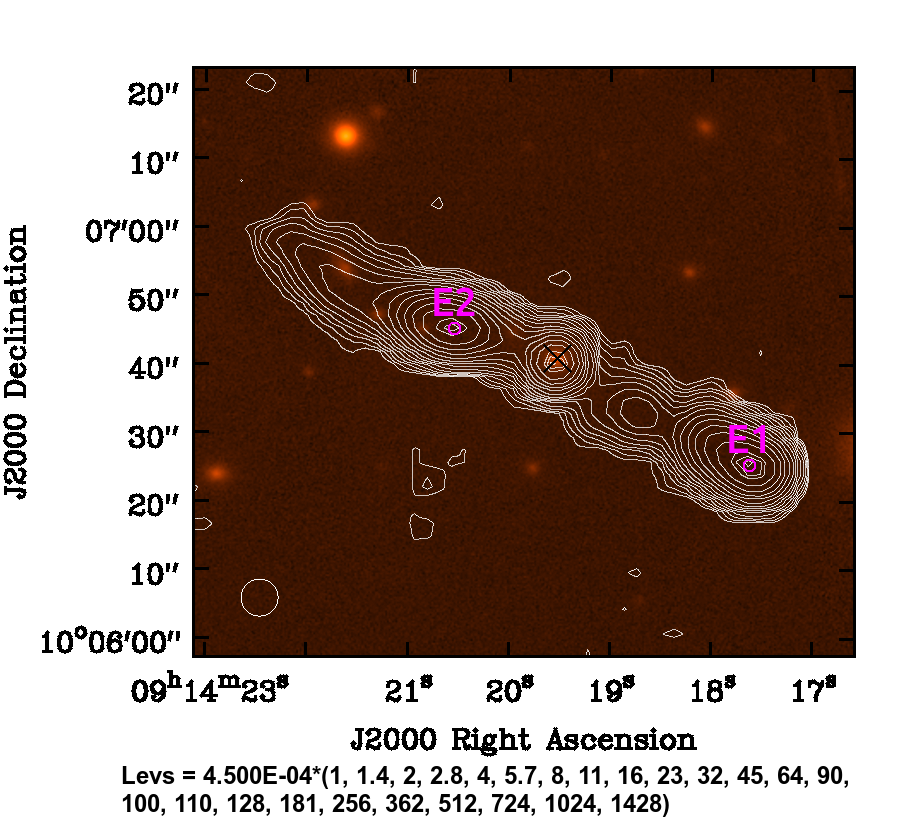}
\includegraphics[width=5.5cm,height=5.5cm,origin=c]{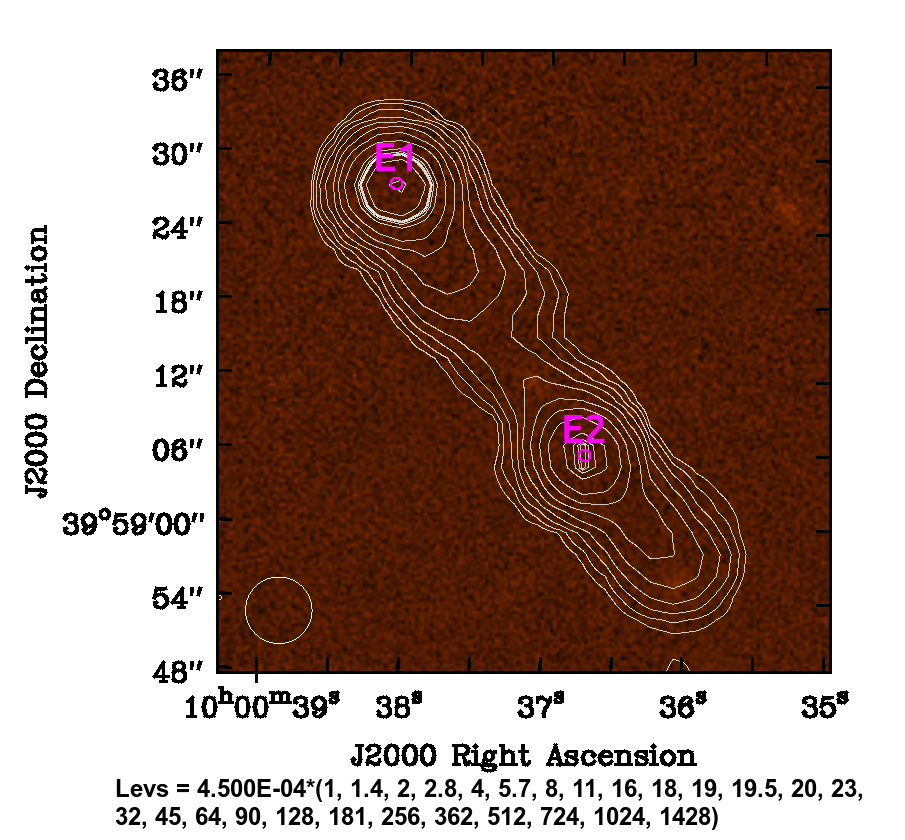}
}
}
\vbox{
\centerline{
\includegraphics[width=5.5cm,height=5.5cm,origin=c]{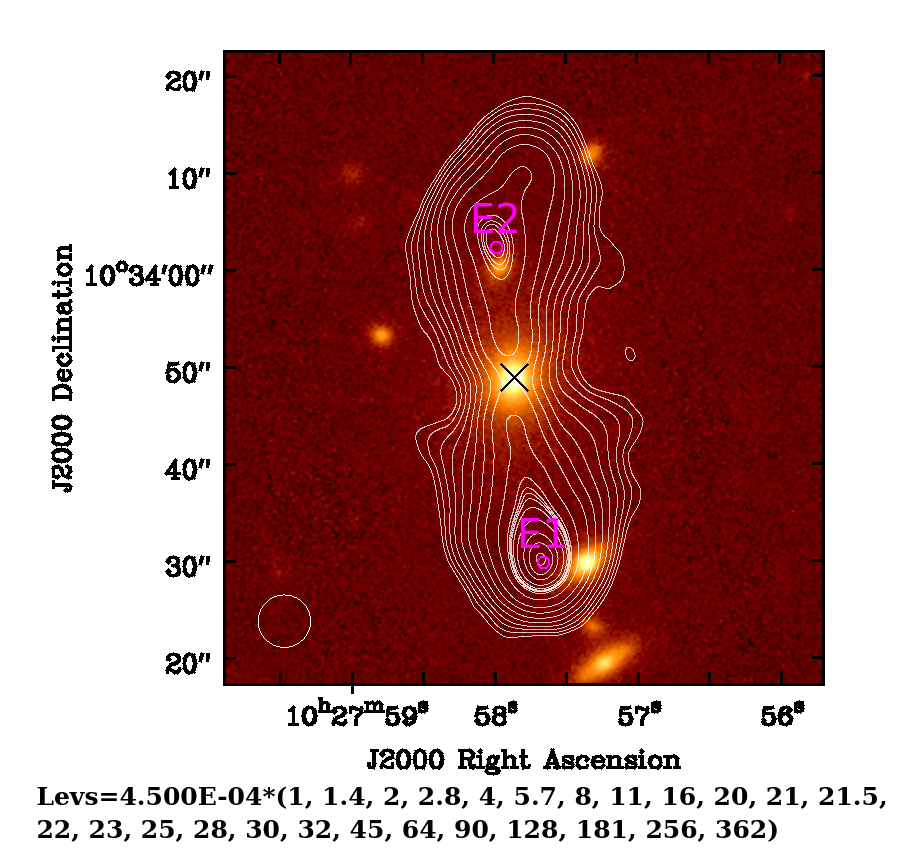}
\includegraphics[width=5.5cm,height=5.5cm,origin=c]{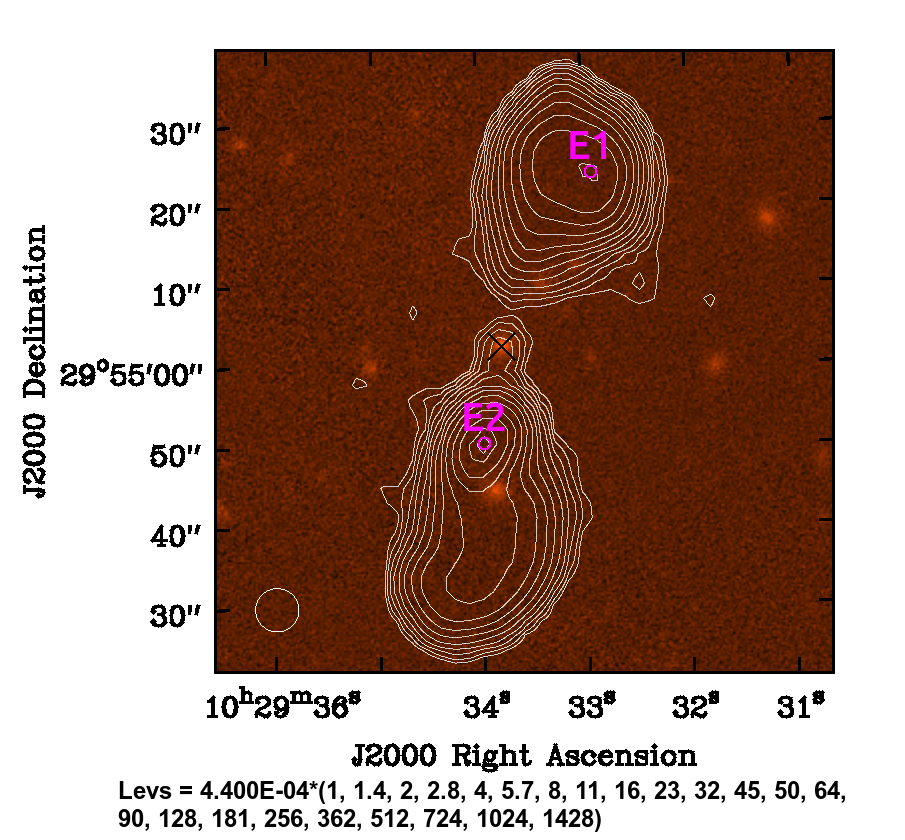}
\includegraphics[width=5.5cm,height=5.5cm,origin=c]{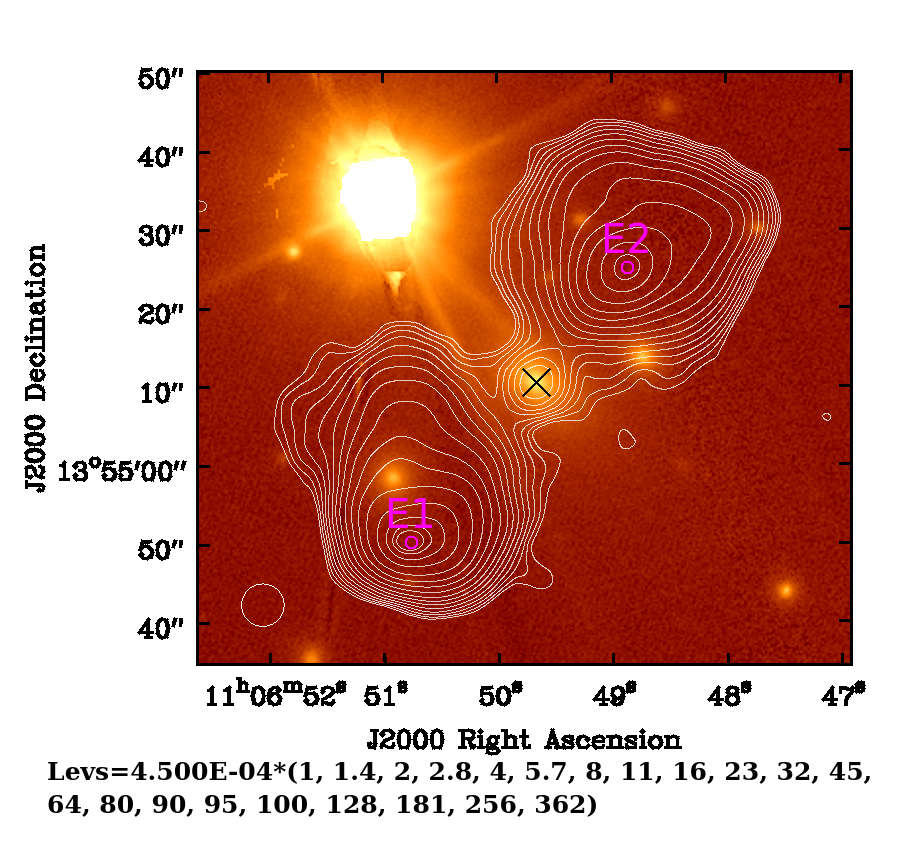}
}
}
\caption{Candidate Hybrid Morphology Radio Sources (HyMoRS) from the FIRST survey \citep{Be95} were shown in contour and corresponding optical images from Pan-STARRS \citep{Ch16} were shown in colour.}
\label{fig:HyMoRS2}
\end{figure*}

\begin{figure*}
\vbox{
\centerline{	
\includegraphics[width=5.5cm,height=5.5cm,origin=c]{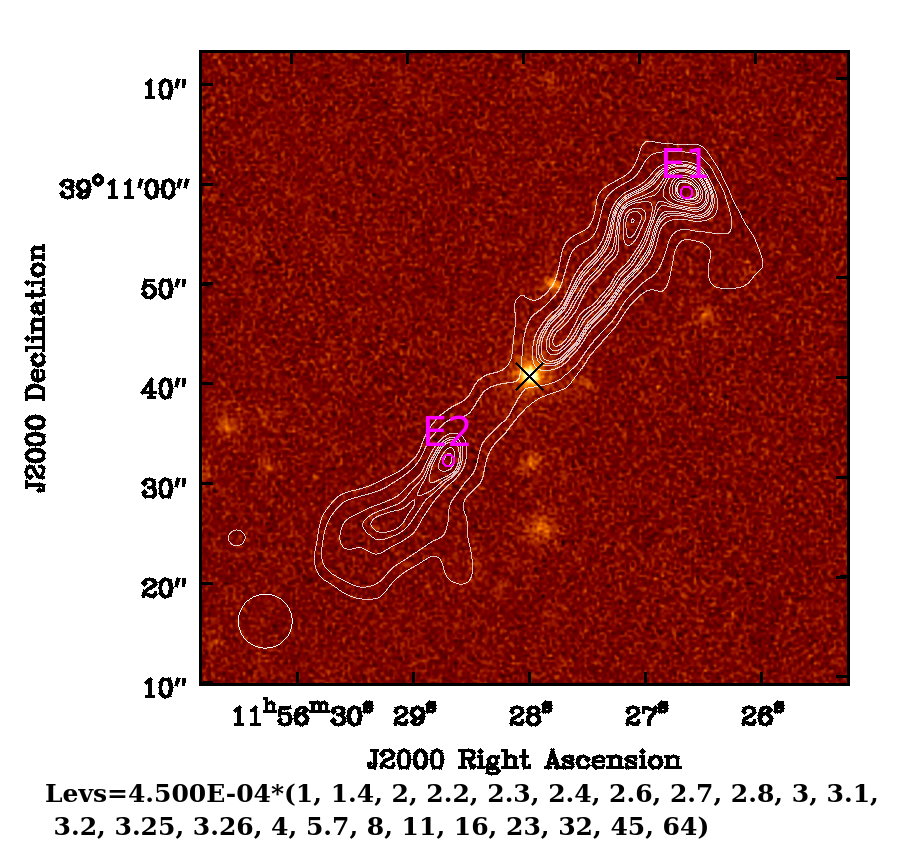}
\includegraphics[width=5.5cm,height=5.5cm,origin=c]{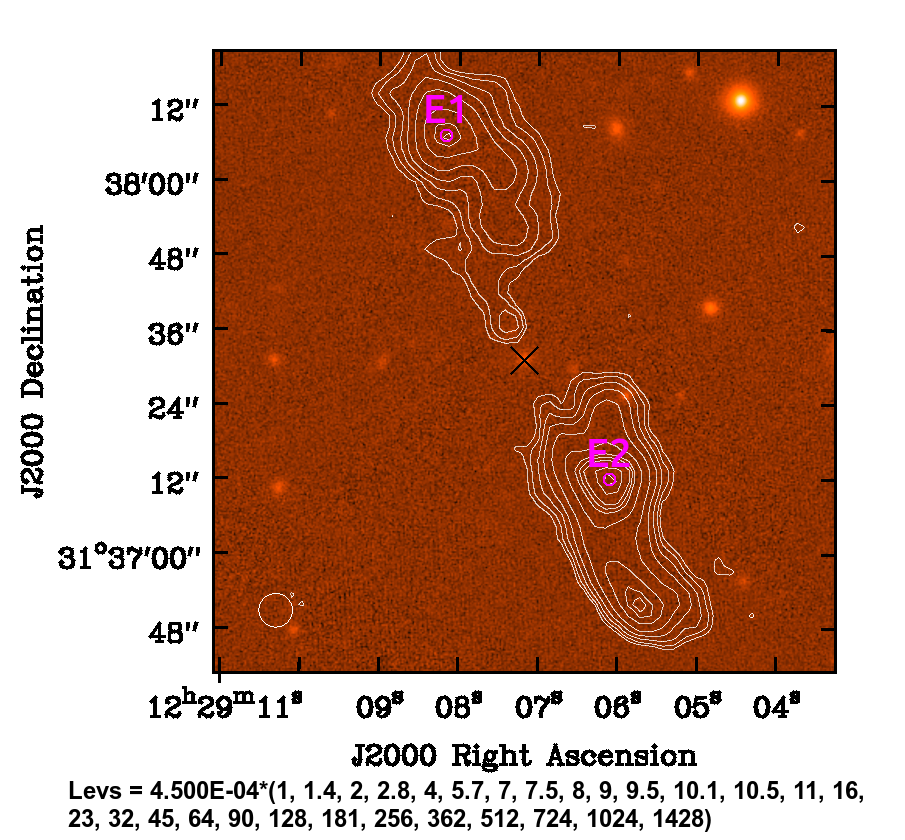}
\includegraphics[width=5.5cm,height=5.5cm,origin=c]{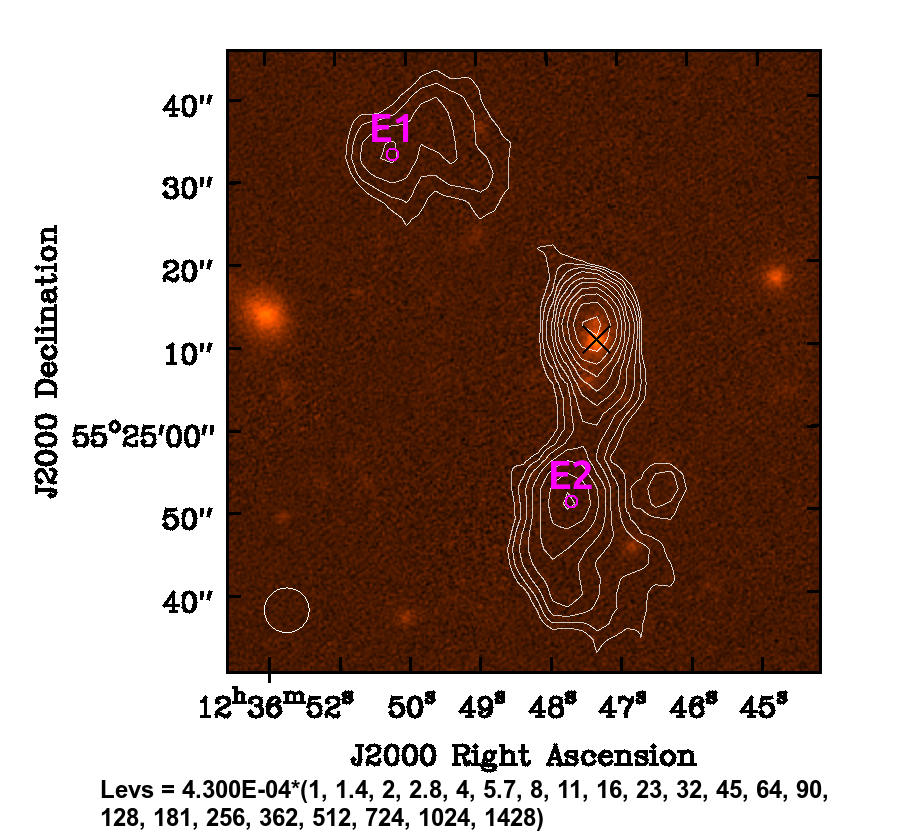}
}
}
\vbox{
\centerline{
\includegraphics[width=5.5cm,height=5.5cm,origin=c]{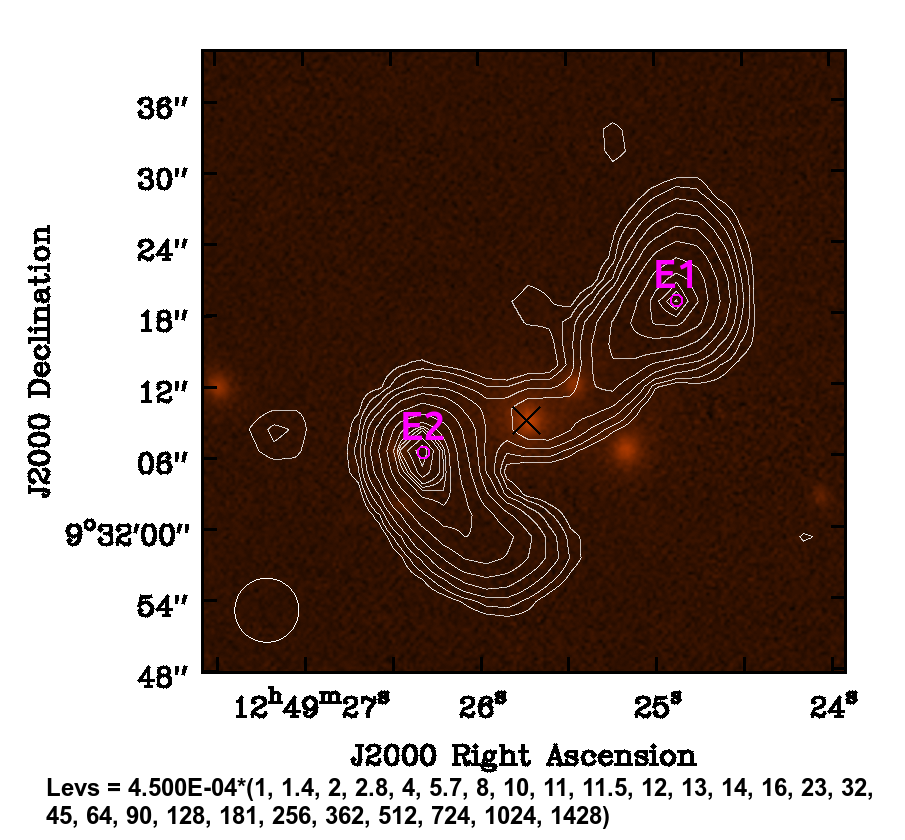}
\includegraphics[width=5.5cm,height=5.5cm,origin=c]{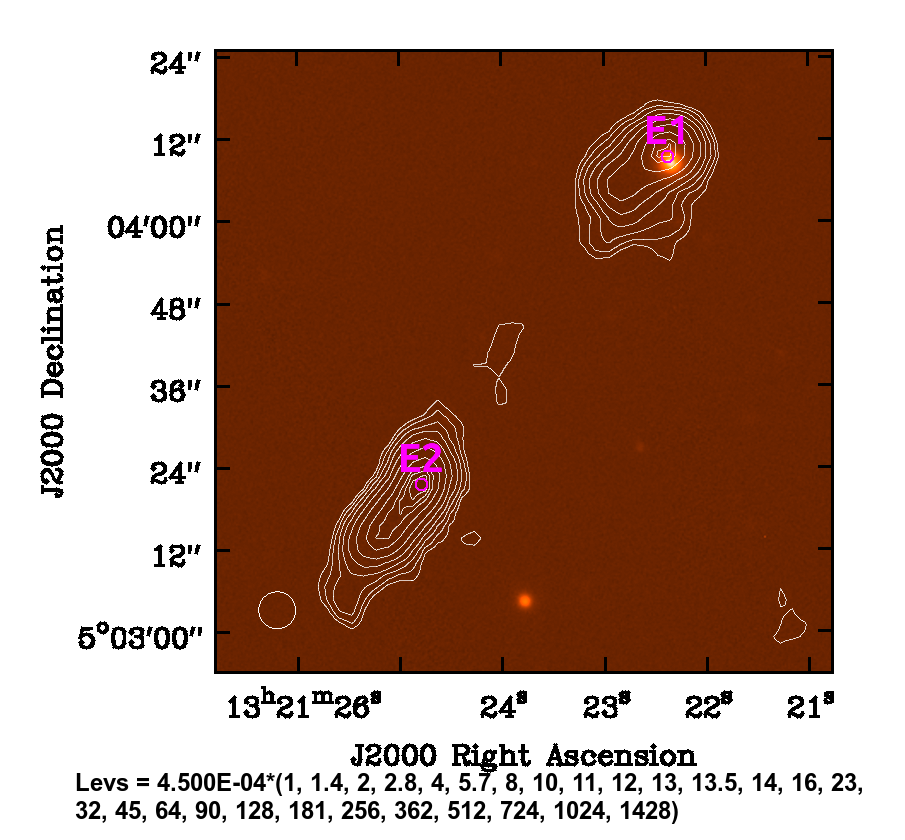}
\includegraphics[width=5.5cm,height=5.5cm,origin=c]{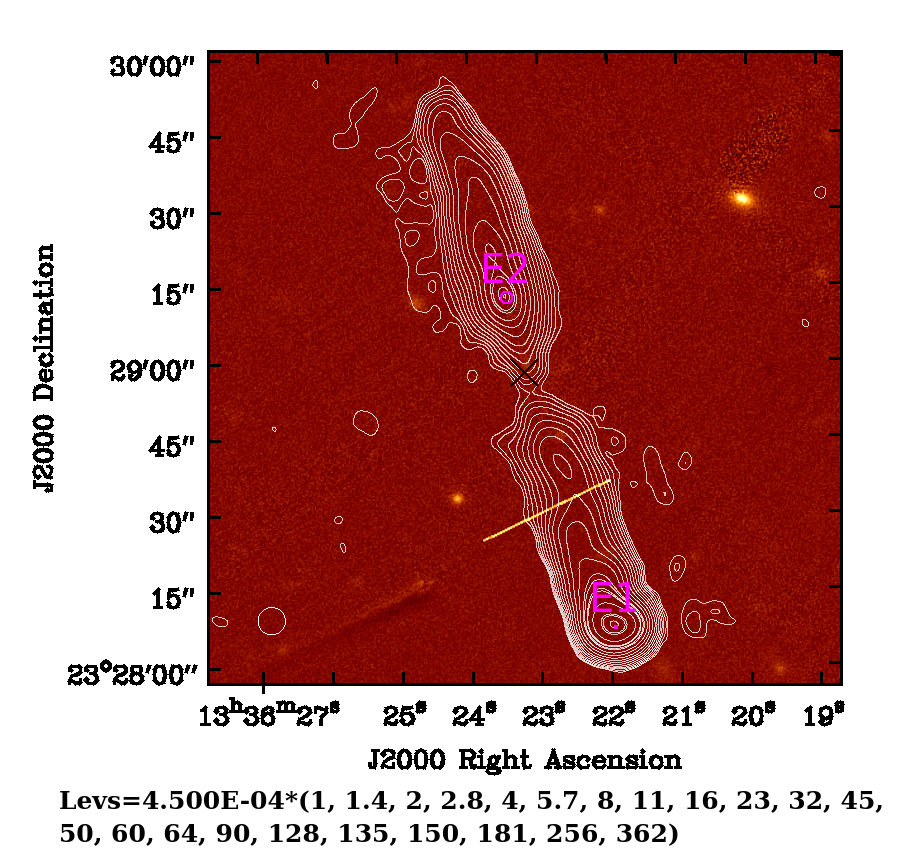}
}
}
\vbox{
\centerline{	
\includegraphics[width=5.5cm,height=5.5cm,origin=c]{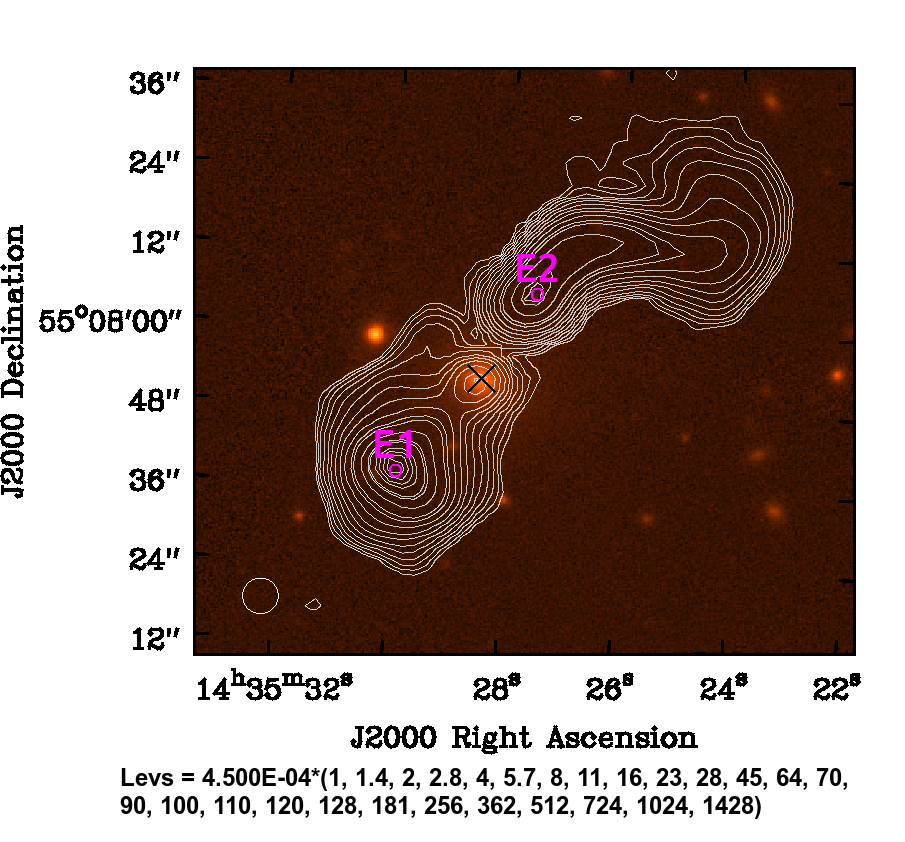}
\includegraphics[width=5.5cm,height=5.5cm,origin=c]{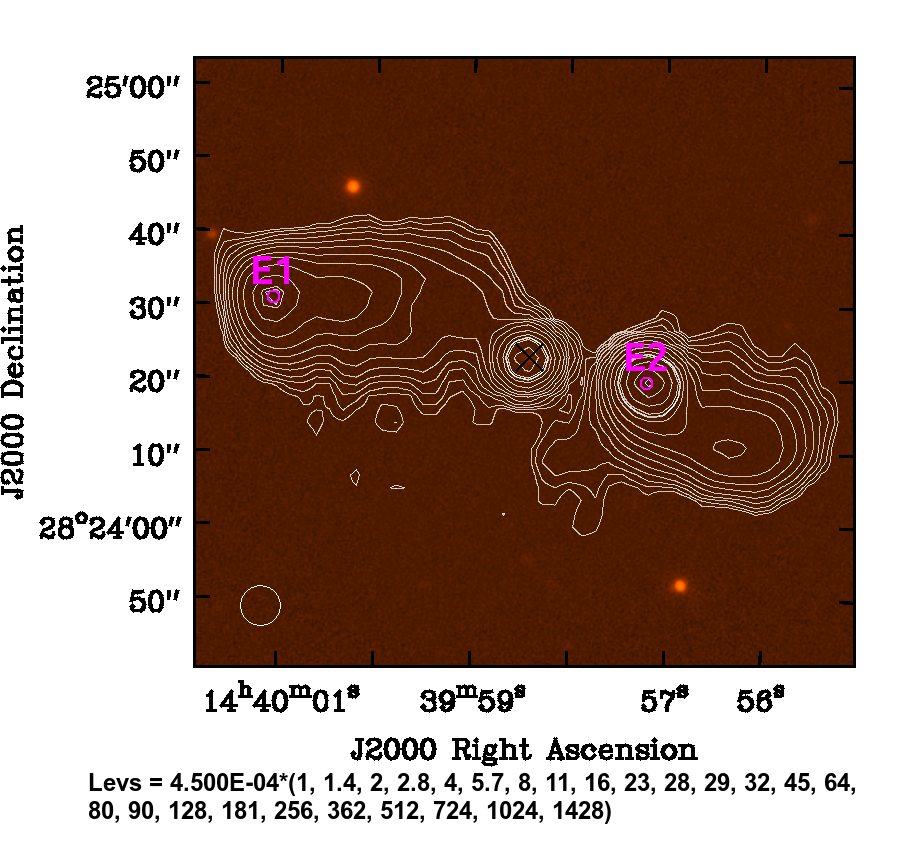} 
\includegraphics[width=5.5cm,height=5.5cm,origin=c]{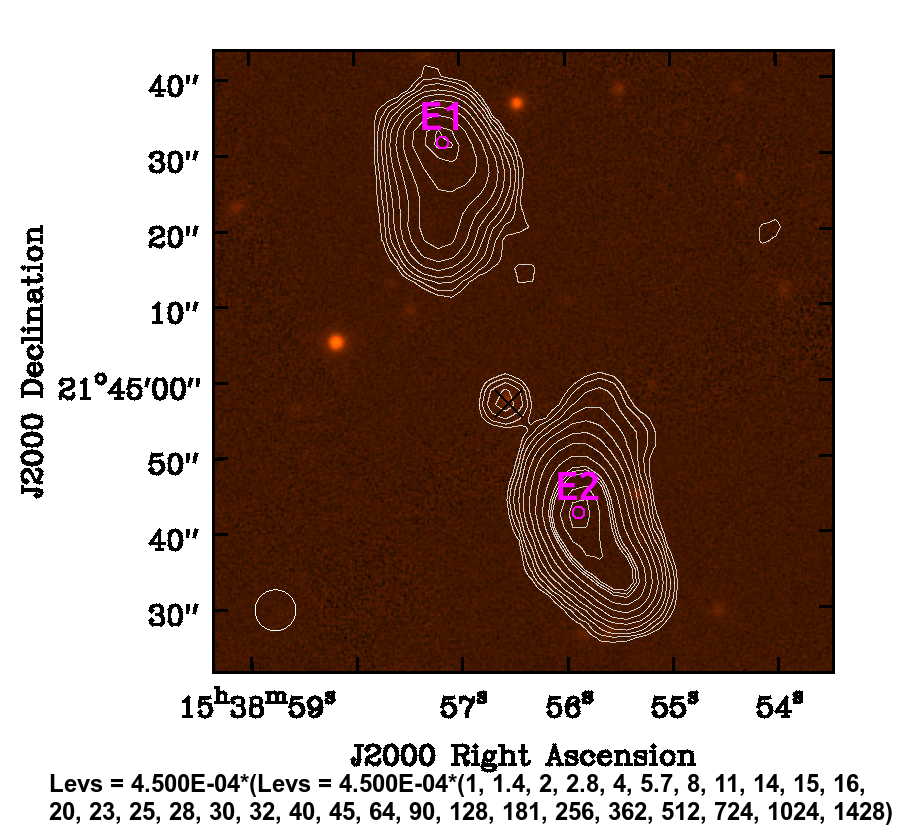}
}
}
\vbox{
\centerline{	
\includegraphics[width=5.5cm,height=5.5cm,origin=c]{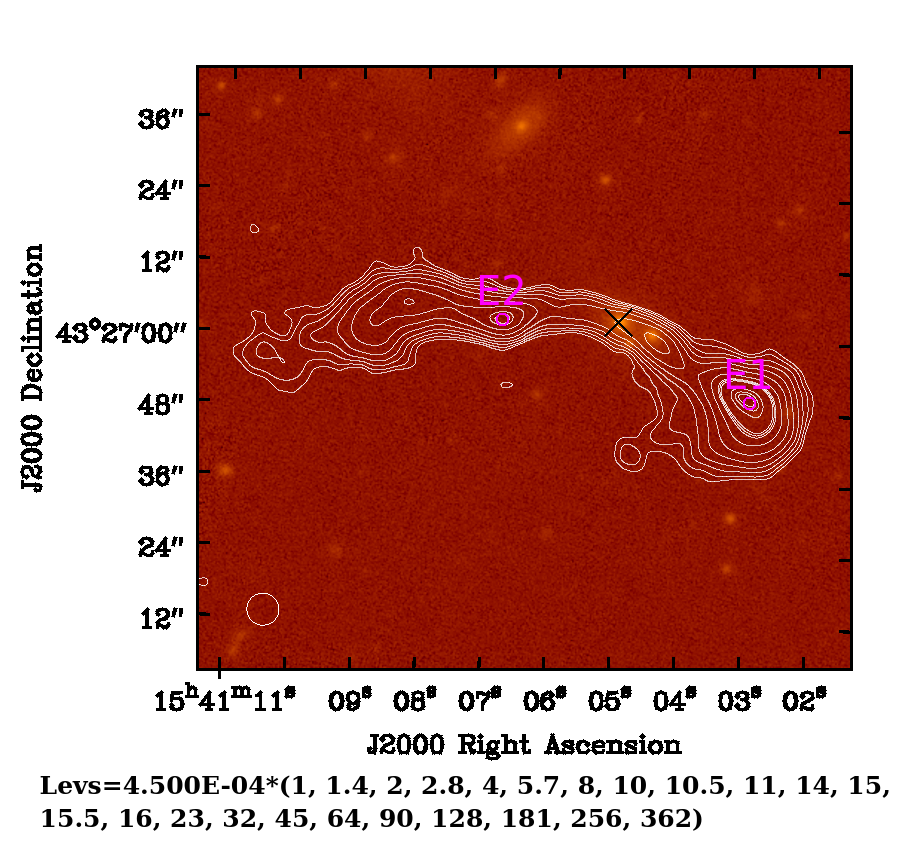}
\includegraphics[width=5.5cm,height=5.5cm,origin=c]{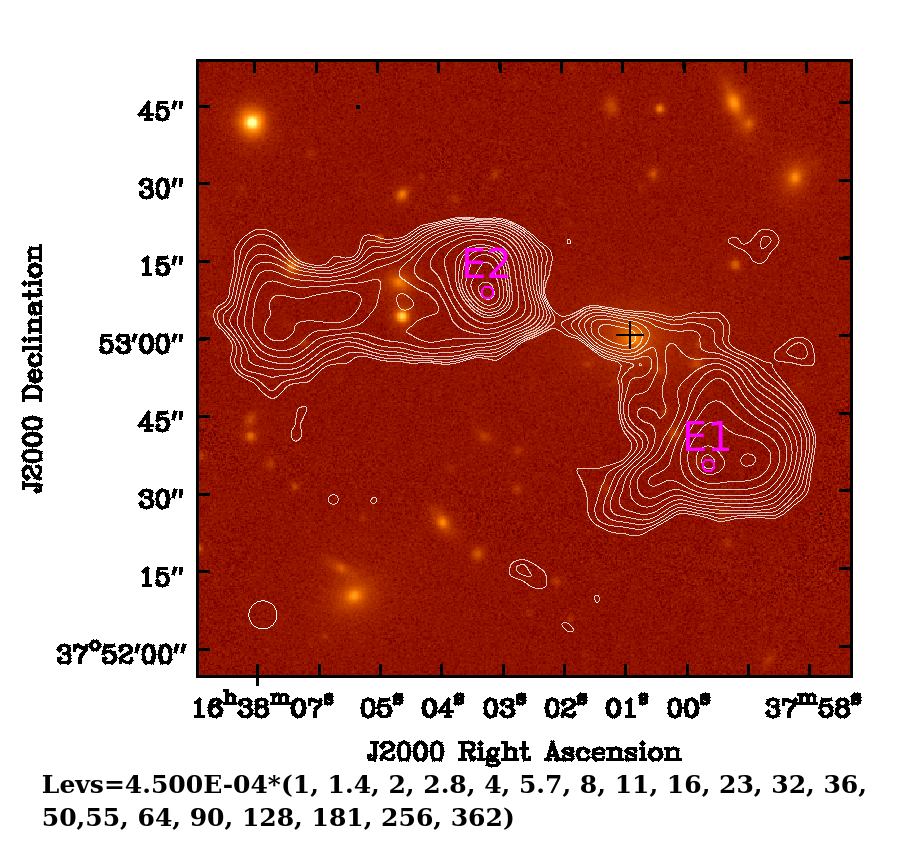}
\includegraphics[width=5.5cm,height=5.5cm,origin=c]{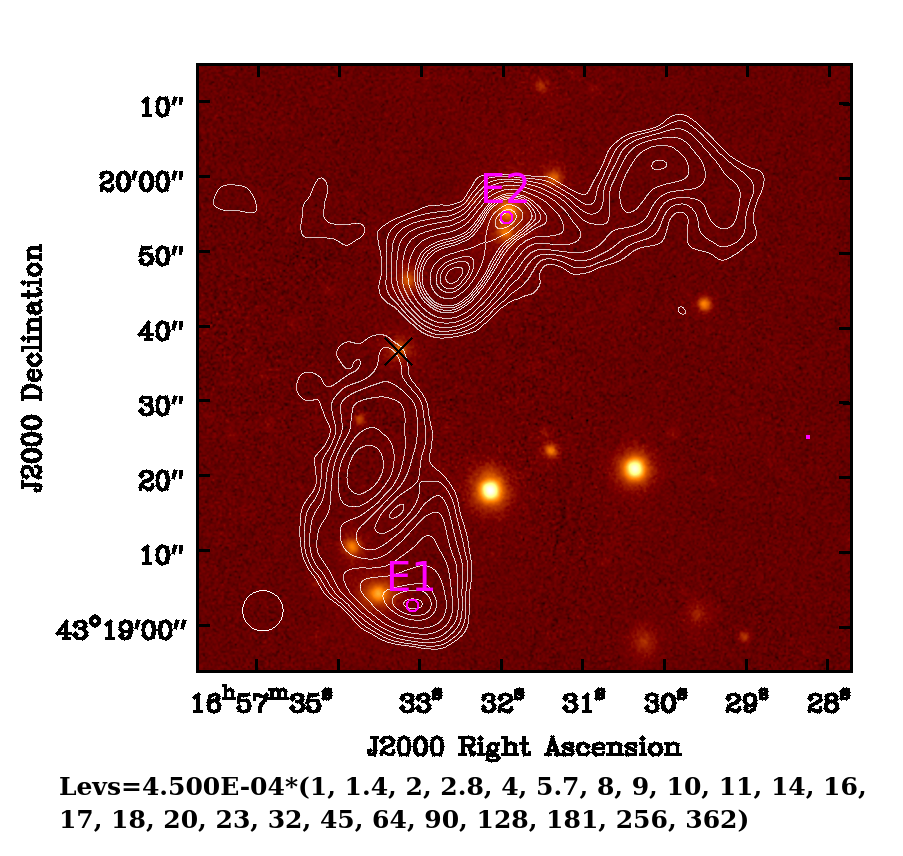}
	}
	}
\contcaption{Candidate Hybrid Morphology Radio Sources (HyMoRS) from the FIRST survey \citep{Be95} were shown in contour and corresponding optical images from Pan-STARRS \citep{Ch16} were shown in colour.}
\end{figure*}

\begin{figure*}
\vbox{
\centerline{
\includegraphics[width=5.5cm,height=5.5cm,origin=c]{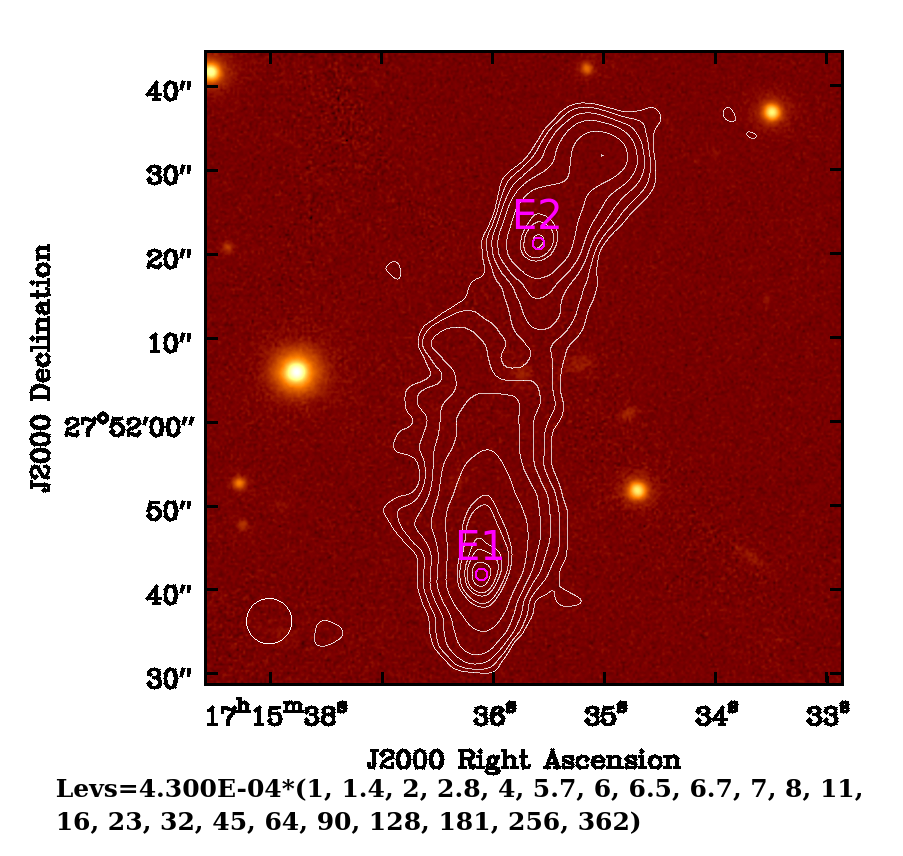}
\includegraphics[width=5.5cm,height=5.5cm,origin=c]{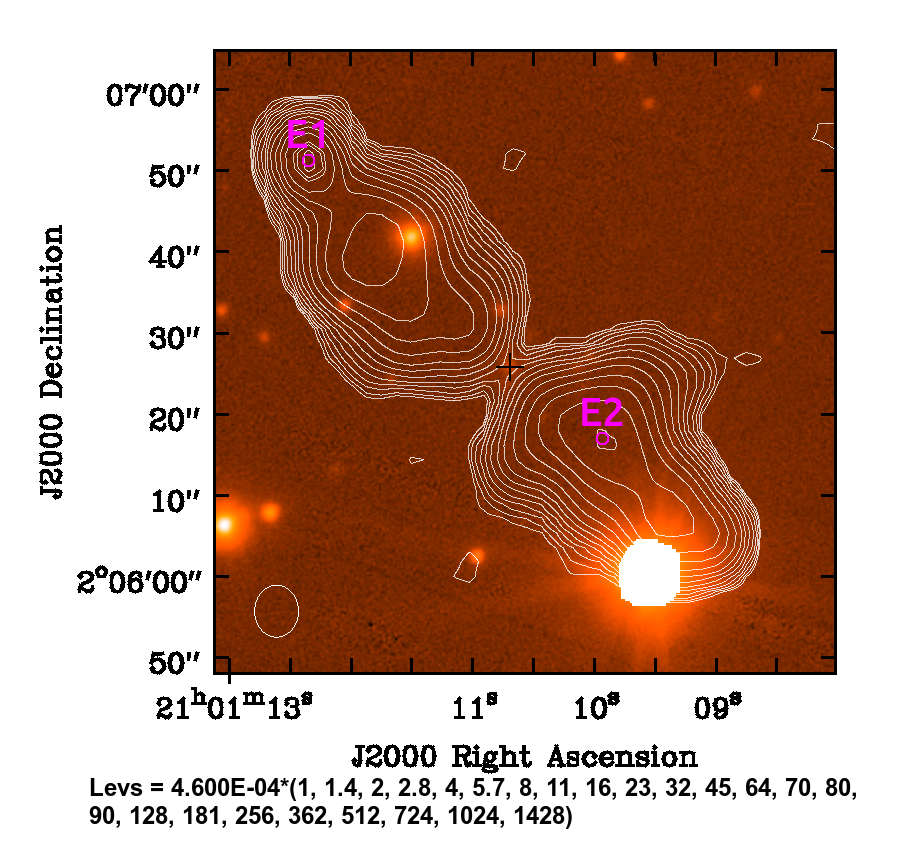}
\includegraphics[width=5.5cm,height=5.5cm,origin=c]{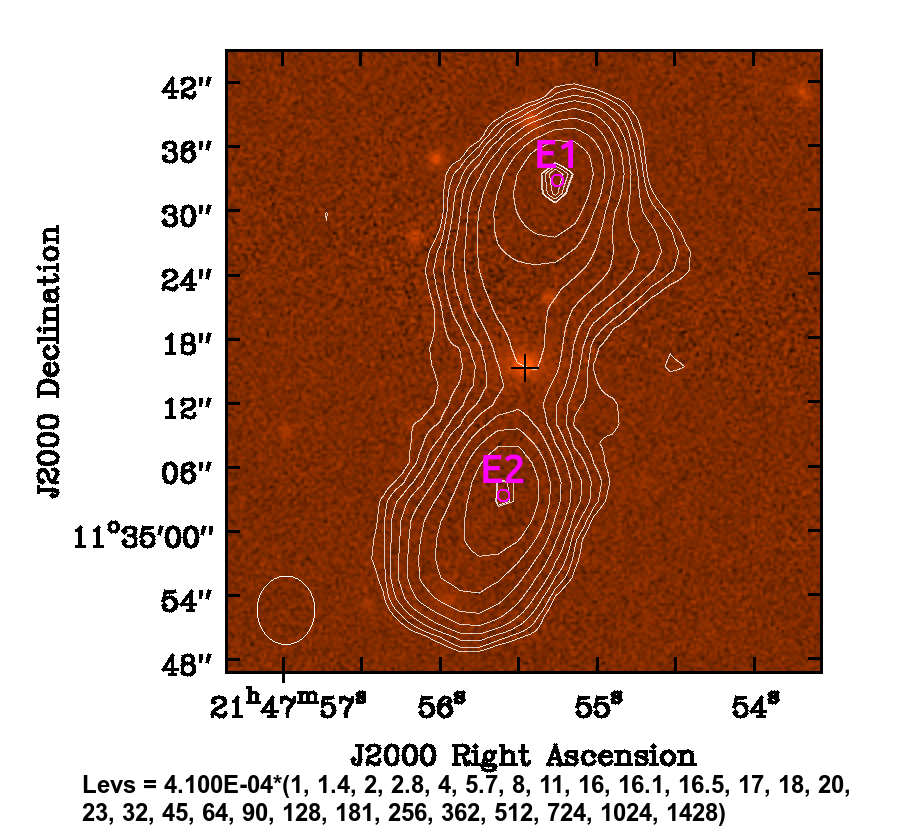}
}
}

\vbox{
\centerline{	
\includegraphics[width=5.5cm,height=5.5cm,origin=c]{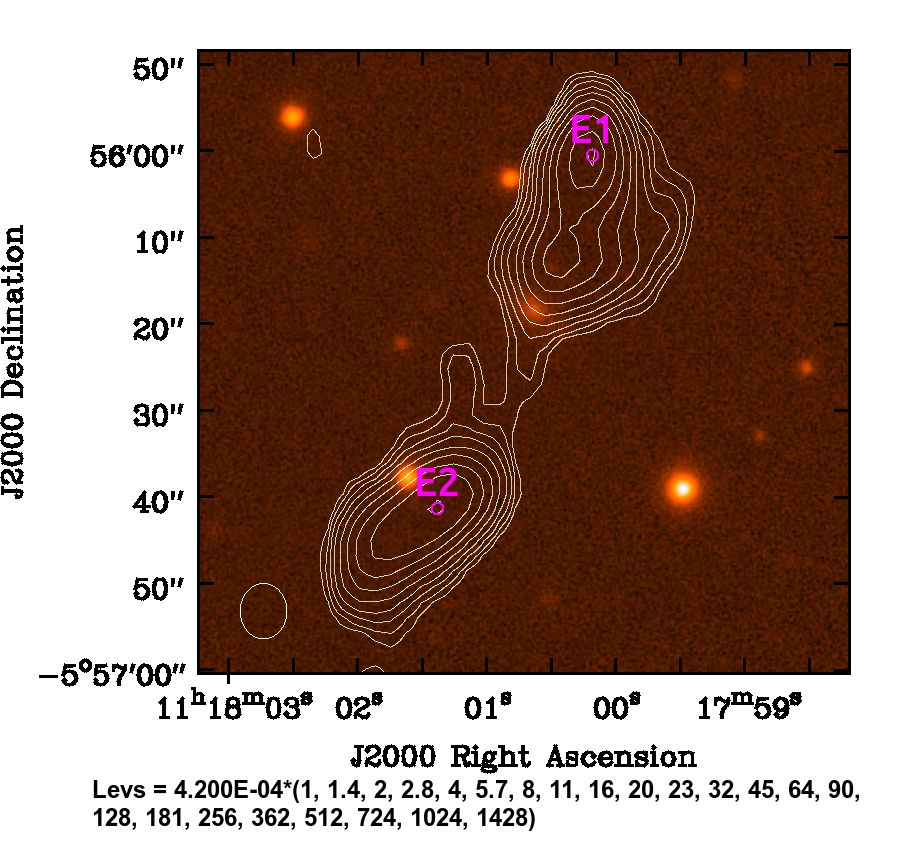}
\includegraphics[width=5.5cm,height=5.5cm,origin=c]{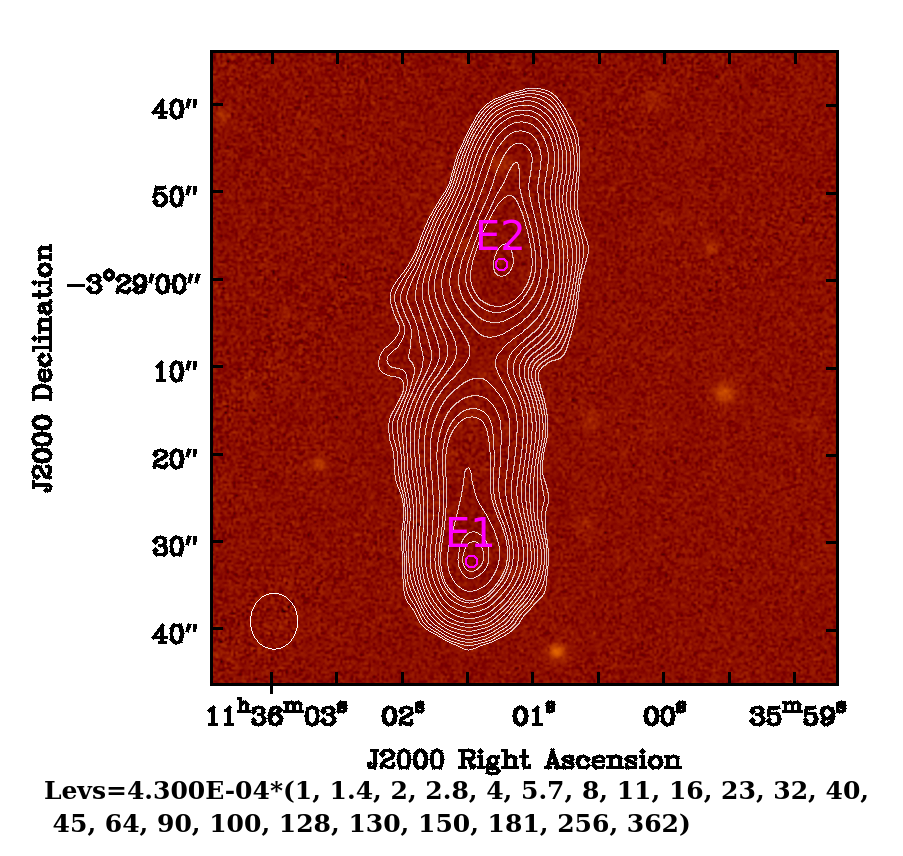}
\includegraphics[width=5.5cm,height=5.5cm,origin=c]{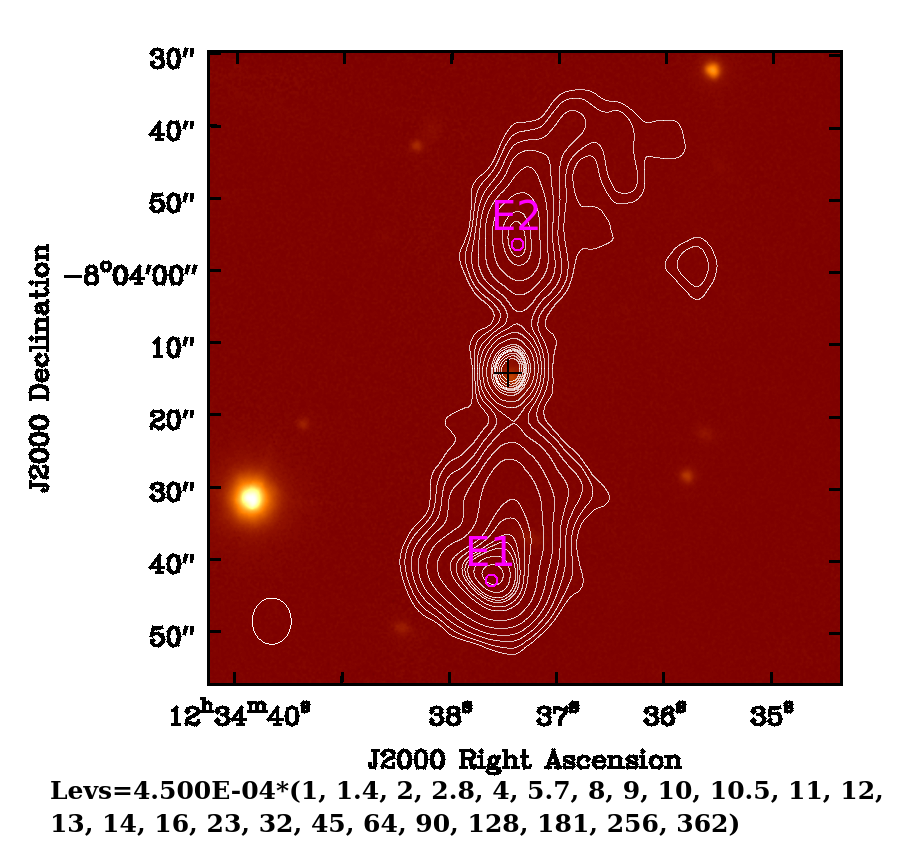}
}
}
\vbox{
\centerline{
\includegraphics[width=5.5cm,height=5.5cm,origin=c]{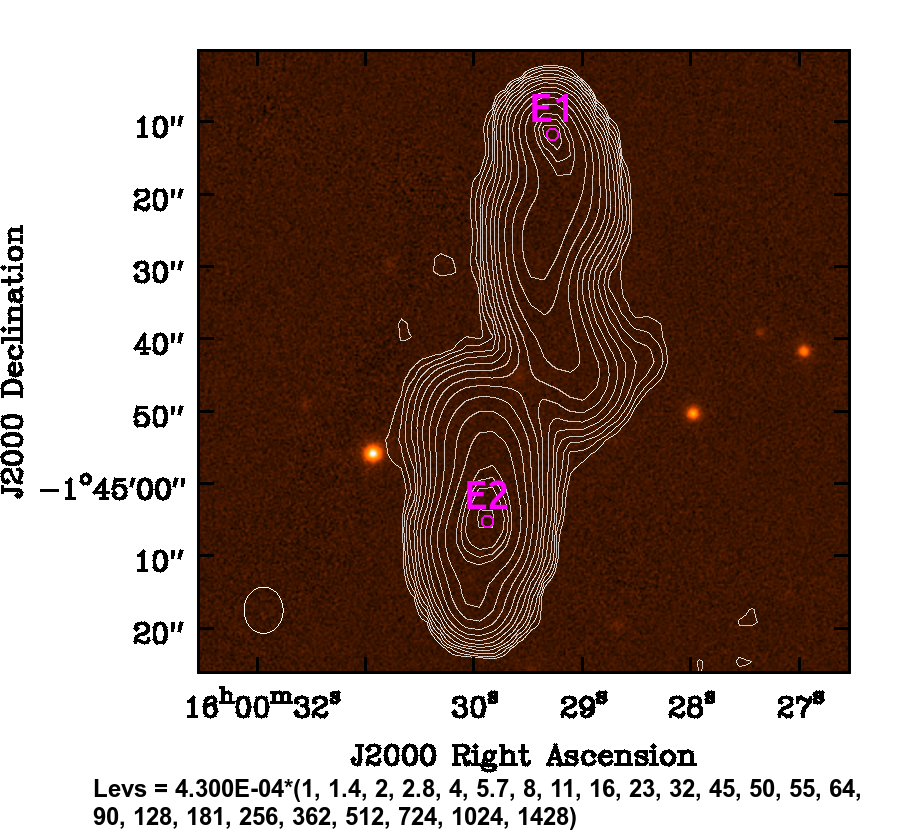}
\includegraphics[width=5.5cm,height=5.5cm,origin=c]{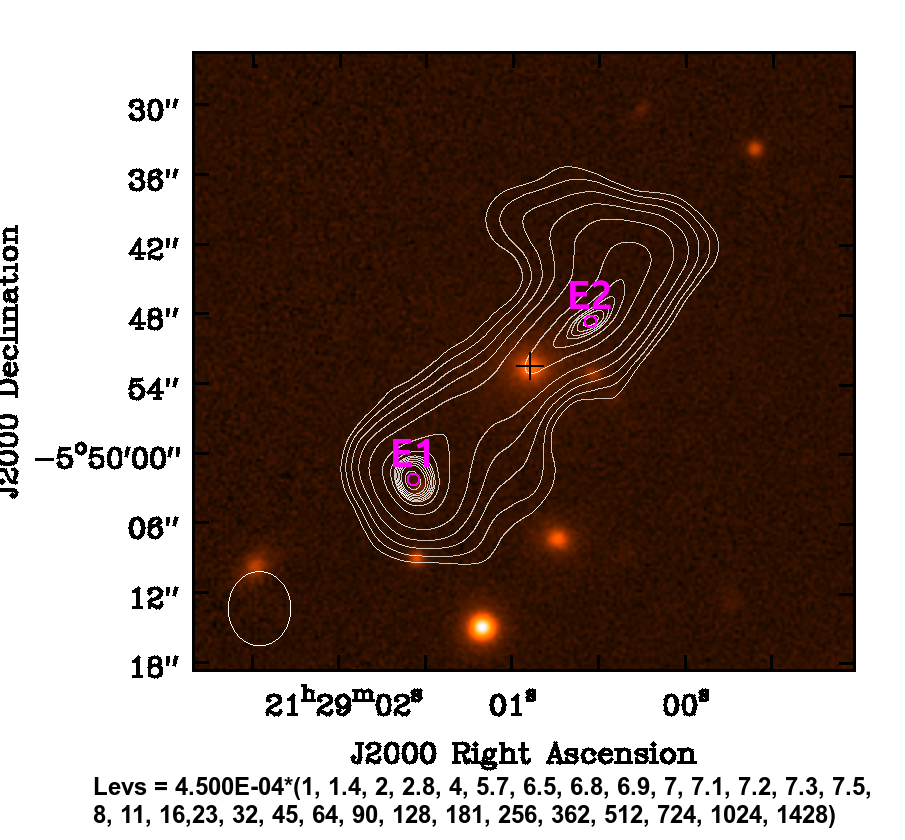}
\includegraphics[width=5.5cm,height=5.5cm,origin=c]{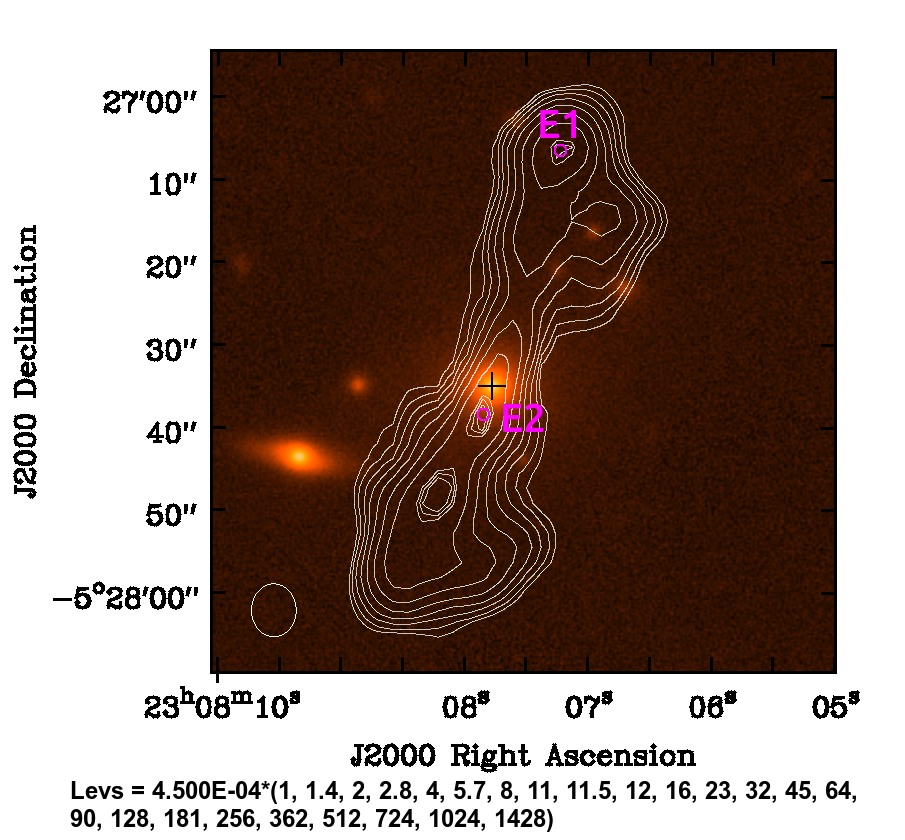}
}
}
\contcaption{Candidate Hybrid Morphology Radio Sources (HyMoRS) from the FIRST survey \citep{Be95} were shown in contour and corresponding optical images from Pan-STARRS \citep{Ch16} were shown in colour.}	
\end{figure*}

For each source in our sample, we measured the corresponding flux density of the source at 150 MHz (column 7) using TGSS \citep{In17}, when the source in TGSS was detected. TGSS suffers from large flux density errors in some regions of the sky. We corrected the flux-density error for the TGSS flux density measurement by using the method described in \citet{Hu17}. We found that corrected values of TGSS flux densities differed slightly from the previous TGSS-only measurement (maximum 7 per cent variation). The two-point spectral index between 1400 and 150 MHz ($\alpha^{1400}_{150}$) was measured, when flux density at 150 MHz was available and tabulated in column 8 of Table \ref{tab:HyMoRS} (for details, see the subsection \ref{subsec:spectral-index}). In column 9, the redshifts ($z$) of these sources were mentioned when available. For sources with a known redshift, the measured luminosity of sources was tabulated in column 10. The common name of sources at radio wavelengths was listed in the last column, which was mentioned earlier without identification of them as HyMoRS. 

The images of all HyMoRS are shown in Fig. \ref{fig:HyMoRS2}. The optical images from Pan-STARRS (optical red filter) were overlaid with the radio images from the VLA FIRST survey. Synthesized beams were shown in the lower-left-hand corner of each image. The location of the optical counterpart of the galaxy was marked with a cross ($\times$) for SDSS counterparts and a plus (${\texttt{+}}$) for Pan-STARRS counterparts in the presented images of HyMoRS. The peaks of FR-II and FR-I flux density in both of the lobes were marked with E1 and E2, respectively, in each image. For each image, the contours were slightly changed to study the morphological structure of each lobe and identify peak points in each lobe. Contour levels of each image were also mentioned.

\begin{table*}
\caption{\bf Hybrid Morphology radio sources}
\label{tab:HyMoRS}
\scriptsize
\begin{tabular}{ccccccccccc} 
\hline
Cat&	Source&R.A.&DEC&  Ref.& $F_{1400}$&$F_{150}$&$\alpha_{150}^{1400}$&$z$  &Luminosity&Common \\
	No.&      name&(J2000.0)&(J2000.0)    &     &(mJy) & (mJy)   & ($\pm$0.05)&& W~Hz$^{-1}$sr$^{-1}$&Name  \\
	   &        &     &               &          &           &        &   &  &$\times10^{25}$&\\	
\hline
	~1     &       J0201+0833          &02:01:23.52 &      +08:33:13.5      &Pan-STARRS&~328$\pm16.40$&~1542$\pm154.2$&0.69&--&--&PMN J0201+0832 \\
	~2	   &	   J0257+0638          &02:57:10.81 &      +06:38:03.8    &Pan-STARRS&1624$\pm81.20$&~7930$\pm793.0$&0.71&--&--&PKS J0257+0638\\  
	~3	   &	~~J0715+5528*         &07:15:24.26 &      +55:28:47.6  &Pan-STARRS&~107$\pm5.35$&~~~814$\pm81.4$&0.91&--&--  &--\\  
	~4	   &	~~J0756+3901*  &07:56:27.95 &      +39:01:36.4     &Pan-STARRS&~172$\pm8.60$&~~678$\pm67.8$&0.76&0.45&13.20&--\\  
	~5     &       J0813+0553          &08:13:28.90 &      +05:53:14.3      &--&~136$\pm6.80$&~~736$\pm73.6$&0.76&--&--&--\\
	~6	   &       J0838+3253          &08:38:44.61 &	+32:53:11.8    &SDSS&~114$\pm5.70$&--&--&0.21  &--&7C 083537.29+330346.00\\  
	~7     &       J0855+4911         &08:55:56.37&       +49:11:10.7      &SDSS&~906$\pm45.30$ &~2004$\pm200.4$&0.35&0.09&~1.86 &--\\
	~8     &       J0914+1006          &09:14:19.52 &      +10:06:40.9      &SDSS&~479$\pm23.95$&~1906$\pm190.6$&0.62&0.31&14.40&MRC 0911+103\\
	~9	   &	   J1000+3959          &10:00:37.09 &      +39:59:11.5     &--&~~55$\pm2.75$&~~285$\pm28.5$&0.74&--&--         &--  \\
	10	   &	   ~J1027+1033$^b$          &10:27:57.87 &	+10:33:48.6     &SDSS&~138$\pm6.90$&~~437$\pm43.7$&0.52&0.11&~0.44&--\\
	11	   &	   J1029+2954          &10:29:33.80 &      +29:55:02.2     &SDSS&~253$\pm12.65$&~1305$\pm130.5$&0.73&0.46&19.22&--\\  
	12	   &	~~J1106+1355$^\alpha$ &11:06:49.67 &	+13:55:10.3  &SDSS&~555$\pm27.75$&~2284$\pm228.4$&0.63&0.12&~2.14&--\\ 
    13	   &	   J1156+3910          &11:56:27.98 &      +39:10:40.3    &SDSS&~~27$\pm1.35$&~~~88$\pm8.8$&0.53&0.42&~1.56&--\\  
	14	   &	   J1229+3137          &12:29:07.17 &      +31:37:31.1   &SDSS&~137$\pm6.85$ &~1009$\pm100.9$&0.89 &0.50&13.48&4C 31.41A\\   
	15     &       J1236+5524          &12:36:47.33 &+55:25:10.5      &SDSS&~~59$\pm2.95$&~~182$\pm18.2$&0.50&0.32&~1.81&LEDA 2501415\\ 
	16     &       J1249+0932          &12:49:25.73 &      +09:32:09.0      &SDSS&~~57$\pm2.85$&~~192$\pm19.2$&0.54&0.23&~0.88&--\\
	17     &       J1321+0503          &13:21:24.22 &      +05:03:43.3      &--&~~61$\pm3.05$&~~340$\pm34.0$&0.77&--&--&-- \\
	18	   &	   J1336+2329          &13:36:23.20 &      +23:28:58.9      &SDSS&~534$\pm26.70$&~4192$\pm428.2$&0.93 &0.61&85.44&--\\ 
	19     &       J1435+5508          &14:35:28.46&       +55:07:52.1      &SDSS&~482$\pm24.10$ &~3323$\pm332.3$&0.86&0.14 &~2.61&--\\  
	20     &       J1439+2824          &14:39:58.42 &      +28:24:22.6      &SDSS&~305$\pm15.25$&~1446$\pm144.6$&0.70&0.36&13.10&--\\
	21	   &	~~J1538+2144*         &15:38:56.56 &	+21:44:57.4     &SDSS&~114$\pm5.70$&~~505$\pm50.5$&0.67&0.47&~8.86&--\\
	22     &       J1541+4327          &15:41:04.95&       +43:27:02.8     &SDSS&~123$\pm6.15$&~~619$\pm61.9$&0.72&0.27&~2.74&LEDA 2221056\\     
	23	   &	   ~J1638+3753$^b$          &16:38:00.93 &      +37:53:00.2   &Pan-STARRS&~654 $\pm32.70$&~2617$\pm261.7$&0.62  &0.16&~4.70&4C +37.48\\
	24     &	   J1657+4319          &16:57:31.94 &	+43:19:55.6    &SDSS&~104$\pm5.20$&~~436$\pm43.6$&0.64&0.20  &~1.21&LEDA 2218211\\                                   
	25	   &	   J1715+2751          &17:15:35.68 &      +27:52:10.1    &--&~~80 $\pm4.00$&~~591$\pm59.1$&0.90&--&--&-- \\  
	26	   &	   J2101+0206          &21:01:08.77 &      +02:06:37.1  &Pan-STARRS&~574 $\pm28.70$&~3093$\pm309.3$&0.75&--&--&4C +01.64\\   
	27     &       ~J2147+1135$^b$         &21:47:55.46 &      +11:35:15.6      &Pan-STARRS&~~83$\pm4.15$&--&--&0.36&--&--\\
	28	   &	   J1118--0556         &11:18:00.70 &     --05:56:30.0  &--&~~82$\pm4.10$&~~333$\pm33.3$&0.63&--&--&--\\
	29     &       J1136--0328         &11:36:01.39 &      --03:29:13.2     &--&~296$\pm14.80$&~3022$\pm302.2$ &1.04&0.82&103.06&--\\
	30	   &	   J1234--0804         &12:34:37.42 &      --08:04:14.6    &Pan-STARRS&~~81$\pm4.05$&~~271 $\pm27.1$&0.54&--&--&--\\ 
	31	   &	   J1600--0144         &16:00:29.60 &      --01:44:44.6    &--&~336$\pm16.80$&~2477$\pm247.7$&0.89&--&--&PMN J1600-0144\\
	32	   &	   J2129--0549         &21:29:00.90 &      --05:49:50.9     &Pan-STARRS&~~32$\pm1.60$&--&-- &--&--&--\\
    33	   &    ~~~J2308--0527*$^b$        &23:08:07.77 &      --05:27:35.1 &Pan-STARRS&~120$\pm6.00$&~~293$\pm$29.3&0.40&0.09&~0.25&--\\
\hline
\end{tabular}\\
Notes:$\alpha$ --Quasars;
$b$ --Photometric redshift, except $b$ all were spectroscopic redshift;
* --FR-I flux density higher than FR-II flux density\\

1: NVSS \citep{Co98}; 2: VLSS \citep{Co07}; 3: 4C \citep{Pi65, Go67, Ca69}; 4: PMN \citep{Gr94}; 5: PKS \citep{Bo64}; 6: VFK \citep{Va15}; 7: MRC \citep{Ll02}; 8: 2MASS \citep{Hu12}\\
\end{table*}

\subsection{Spectral index ($\alpha_{150}^{1400}$)}
\label{subsec:spectral-index}
The two-point spectral index of newly discovered hybrid morphology radio galaxies between 150 and 1400 MHz was measured assuming $S \propto \nu^{-\alpha}$, where $\alpha$ is the spectral index and $S$ is the radiative flux density at a given frequency $\nu$. In Table \ref{tab:HyMoRS}, spectral index ($\alpha_{150}^{1400}$) was mentioned for 30 HyMoRS. The remaining three HyMoRS were not detectable in TGSS due to low flux density.

\begin{figure}
\centerline{
\includegraphics[height=7cm,width=8cm,origin=c]{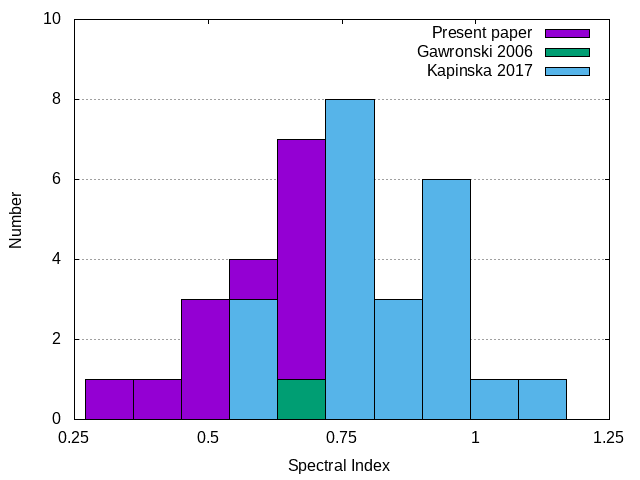}
}
\caption{{Histogram showing spectral index distribution of HyMoRS. Here purple color represented HyMoRS in present paper and blue color represented HyMoRS reported previously in \citet{Ka17} and green color represented HyMoRS detected by \citet{Ga06}.}}
\label{fig:Spectral-index}	
\end{figure}

In Fig. \ref{fig:Spectral-index}, a histogram with spectral index distribution for all HyMoRS presented in this paper was shown. We also included HyMoRS detected by \citet{Ka17} and \citet{Ga06} in the figure. The histogram peaked near 0.75 for all located HyMoRS. The average spectral index for five bright and extended HyMoRS detected by \citet{Ga06} was $\alpha_{150}^{1400}$ =0.83, which was near the peak of this histogram. In Fig. \ref{fig:Spectral-index}, it was seen that the peak of spectral index of HyMoRS detected by \citet{Ka17} was slightly high (average of spectral index by \citet{Ka17}  $\alpha_{150}^{1400}$ =0.82) in comparison to this paper and HyMoRS detected by \citet{Ga06}. The steeper spectral index in \citet{Ka17} was due to a bias in source selection as it preferred brighter and extended sources compared to this paper. The total span of $\alpha_{150}^{1400}$ of sources presented in this paper ranged from $0.35$ to $1.04$. Most of the hybrid morphology radio galaxies (93.3 per cent) showed a steep radio spectrum $\alpha\ge0.5$. 

The uncertainty of spectral index measurements between NVSS (1400 MHz) and TGSS (150 MHz) frequencies due to flux density uncertainty \citep{Ma16} is
\begin{equation}
	\Delta\alpha=\frac{1}{\ln\frac{\nu_{1}}{\nu_{2}}}\sqrt{\left(\frac{\Delta S_{1}}{S_{1}}\right)^{2}+\left(\frac{\Delta S_{2}}{S_{2}}\right)^{2}}
	\label{equ:equ2}
\end{equation}
where $\nu_{1, 2}$ and $S_{1, 2}$ refer to NVSS and TGSS frequencies and flux densities respectively. The flux density accuracy in NVSS and TGSS ADR1 is $\sim5$ per cent \citep{Co98} and $\sim10$ per cent \citep{In17}, respectively. The measured spectral index uncertainty between NVSS and TGSS ADR1 measurements using equation \ref{equ:equ2} is $\Delta\alpha$=0.05. In both HyMoRS J2308--0527 and J0855+4911, a flat spectrum $(\alpha<0.5)$ was observed.

Among all the HyMoRS, J1136--0328 had the highest spectral index (with $\alpha^{1400}_{150} = 1.04$) and J0855+4911 had the lowest spectral index (with $\alpha^{1400}_{150} = 0.35$). The mean and median spectral index, presented in this paper, were 0.70 and 0.70, respectively (with 1$\sigma$ standard deviation = 0.16), which were nearly equal to the spectral index of normal radio galaxies \citep{Ku06, Ku10, Da02a, Da02b}.
	  
\subsection{Radio luminosity}
\label{subsec:lum}

\begin{figure}	
\centerline{
\includegraphics[height=7cm,width=8cm,angle=0,origin=c]{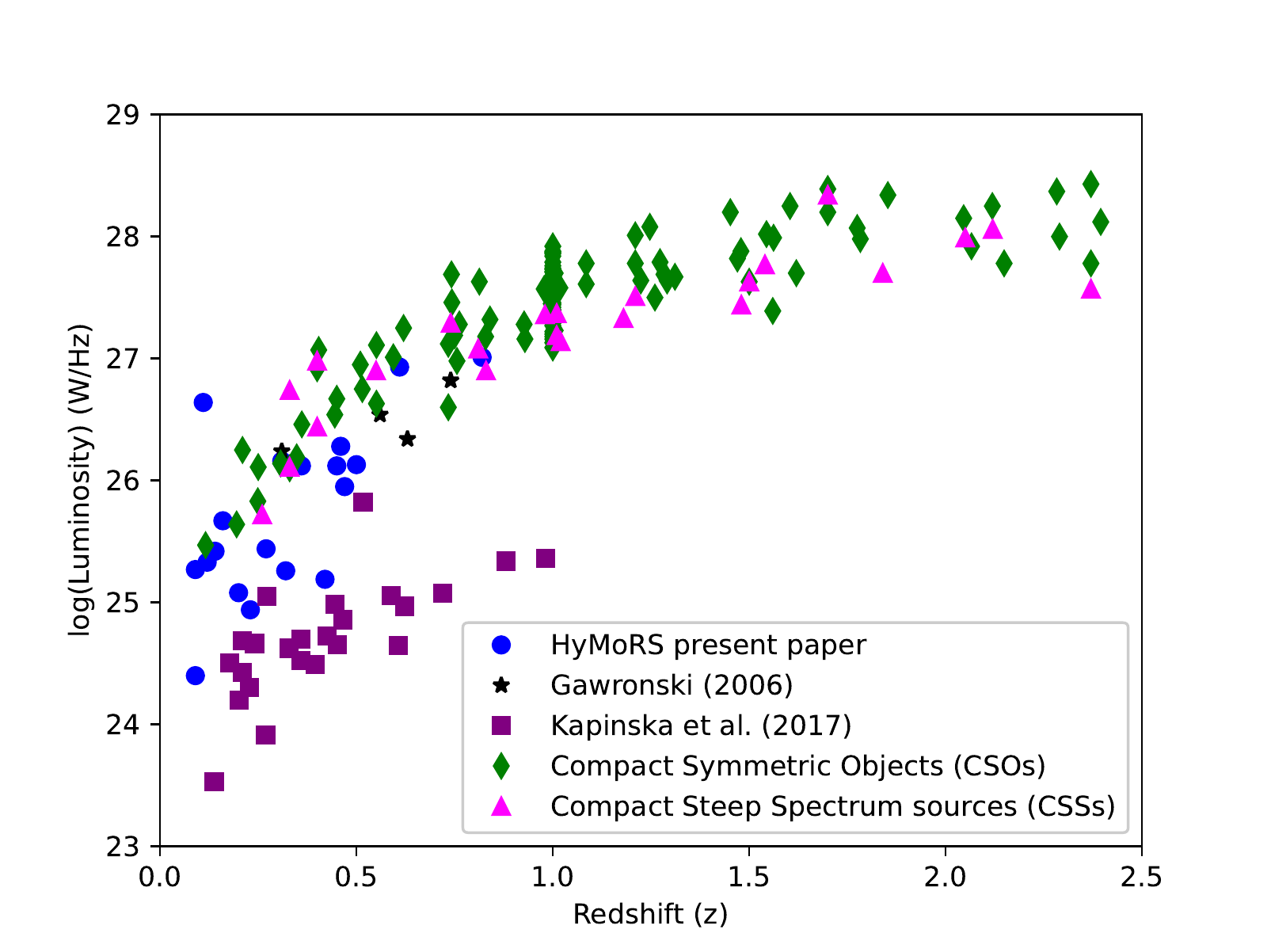}
}
\caption{Distribution of the radio luminosity ($\log L_{\textrm{rad}}$) with the redshift ($z$) for all HyMoRS presented in this paper was indicated with a blue circle. For seventeen sources, measured redshifts were spectroscopic, and for four sources, measured redshifts were photometric. Four previously detected HyMoRS (three photometric and one spectroscopic redshift) by \citet{Ga06, Ce13} were shown in star points with black colour. Square points with purple colour indicated the previously detected HyMoRS by \citet{Ka17}. For comparison, we also included the radio luminosity of the smallest Compact Symmetric Objects (CSOs) \citep{Ku06, Ku10} by green diamond points and low-power Compact Steep-Spectrum sources (CSSs) \citep{Da02a, Da02b} by pink triangle color points.}
\label{fig:redshift-logL}
\end{figure}

We measured the radio luminosity ($L_{\textrm{rad}}$) for all the newly discovered sources (when the value of $z$ was available) using the standard formula \citep{Do09}
\begin{equation}	
    L_{\textrm{rad}}=4\pi{D_{L}}^{2}S_{0}(1+z)^{\alpha-1}
\end{equation}
where $z$ is the redshift of the radio galaxy, $\alpha$ is the spectral index ($S \propto \nu^{-\alpha}$), $D_{L}$ is luminosity distance to the source in metre (m), $S_0$ is the flux density (W m$^{-2}$ Hz$^{-1}$) at a given frequency.

J1136--0328 was the most luminous HyMoRS in our sample, with a $\log L_{\textrm{\textrm{rad}}}=27.01$ W {Hz}$^{-1}$ sr$^{-1}$ and NVSS flux density of 534 mJy. The source J1136--0328 was also the farthest detected HyMoRS ($z=0.82$) in our sample. J2308--0527 was the least luminous HyMoRS in our sample with $\log L_{\textrm{\textrm{rad}}}=24.4$ W {Hz}$^{-1}$ sr$^{-1}$ ($z=0.09$) and an NVSS flux density of 120 mJy.

\section{DISCUSSION}
\label{sec:disc}
In this paper, we listed 33 newly founded candidate HyMoRS from the VLA FIRST survey. We categorized the morphology of these sources as HyMoRS depending on the measured FR index ($R$) on both sides of the radio structures.
 
The previous detection of 25 HyMoRS by \citet{Ka17} was not done systematically as they only listed all HyMoRS found in the citizen science project Radio Galaxy Zoo (RGZ). \citet{Ga06} detected five HyMoRS systematically in which they used filtering criteria and aimed to detect only bright extended HyMoRS inside the five small sub-areas located in the FIRST survey. They missed most of the sources from this paper, maybe because of the selection criteria they chose. In this paper, with the help of the FIRST survey, we visually examined 20 045 (angular size of $>15$ arcsec) sources and successfully detected thirty-three new candidate HyMoRS.

For sources presented in this paper, the range of the FR index for FR-I lobes was between 0.98 and 1.47 and the range of the FR index for FR-II lobes was between 1.69 and 2.31, with an average of 1.28 and 2.04, respectively for FR-I and FR-II lobes. The average value of the FR index for FR-I and FR-II lobes was comparable with 3CRR sources, with an average of 1.13 for FR-I lobes and 2.13 for FR-II lobes \citep{La83}. The average FR index for FR-I lobes and FR-II lobes for HyMoRS presented in \citet{Ka17} was 1.18 and 2.14, which were close to the value of our sample.

The FR dichotomy in radio morphology is linked with the optical brightness of the host via the fundamental difference in parameters such as black hole mass, pressure, and accretion rate, as all of these parameters can affect jet propagation. It is also believed that the supermassive black hole (SMBH) spin can play an important role in the formation of the FR morphology of radio galaxies. SMBH spin with a very high-speed form an FR-II structure and that of a low-speed SMBH spin form an FR-I structure \citep{Go00, Ha20, Me97, Ra02}.

\citet{Bi95} proposed that the FR dichotomy may result from the transition of a supersonic but relatively weak jet to a subsonic or transonic flow near the innermost ($\sim$1 kpc) region of the host giant galaxy due to deceleration by thermal plasma. There are more fundamental differences between FR-I and FR-II classes due to the composition of jets. For FR-I sources, e$^-$-- e$^+$ plasma is preferred in the structure of jets, while in the case of FR-II sources, e$^-$-- p could be preferred \citep{Re96}. Through spectroscopic analysis of emission line nebulae associated with radio galaxies, \citet{Ba92, Ba95} proposed that the angular momentum of the accretion disc may also be important in forming the FR morphologies of extragalactic radio sources. 

Recently, the discovery of low-luminosity FR-II sources in the LOFAR Two-meter Sky Survey (LoTSS) has put into question the classical FR definition based on radio luminosity \citep{Mi19}. There are no statistically significant differences found in terms of radio luminosity between FR-I High-Excitation Radio Galaxies (HERGs) that are very bright in optical and FR-II Low-Excitation Radio Galaxies (LERGs) that are generally found in fainter hosts) \citep{Ca17, Sh17, Ha17, Mi19}. It is suggested that in terms of radio luminosity, the separation of the two morphological populations in the FR dichotomy is not as clear-cut as initially reported by \citet{Le96}. HyMoRS may be a transient phenomenon that requires central engine modulation and differential light travel time between the approaching and receding regions of the radio source to generate this morphology \citep{Go96, Ha20}.

Out of 33 sources reported in this paper, redshift was known for 21 HyMoRS, of which seventeen had spectroscopic redshift and four had photometric redshift. One source, J1106+1355, in our sample had quasar like behaviour \citep{Sh15}. 

Amongst the HyMoRS candidates presented in this paper, sources like J0855+4911, J1638+3752, J1439+2824, J1156+3910, J1657+4319, J0201+0833, and J0756+3901 had multiple hotspots in the structure. For these sources, we calculated the FR-index based on the location of the brightest hotspot on each side of the structure. The southern part of the source J1234--0804 had an FR-II like structure with a bright hot spot at the extreme edge, and the north-eastern part had a diffused FR-I like structure. HyMoRS with multiple hotspots was also found previously in \citet{Pi11}. It was postulated that this observed morphological asymmetry with multiple hotspots might be due to the re-started activity in some of the radio galaxies \citep{Ma12a, Ma12b}.
Sources with multiple hotspots are not uncommon in radio galaxies, and these sources are thought to be transitional FR-IIs, either dying or restarting \citep{Li10}. A thorough investigation of the properties of these sources is required \citep{Sa12}.  

For sources presented in this paper, radio luminosities at 1400 MHz were in the order of $10^{25}$ W Hz$^{-1}$sr$^{-1}$, which was similar to a typical radio galaxy \citep{Fa74}. For FR-I radio galaxies, $L_{\textrm{rad}}<2\times10^{25}$ ($\log L_{\textrm{rad}}=25.3$) W Hz$^{-1}$sr$^{-1}$ and for FR-II, $L_{\textrm{rad}} > 2\times10^{25}$ ($\log L_{\textrm{rad}}=25.3$) W Hz$^{-1}$sr$^{-1}$ \citep{Fa74}. The sources presented in this paper were more luminous than the previously detected twenty-five candidates HyMoRS by \citet{Ka17} and less luminous than the five detected HyMoRS by \citet{Ga06, Ce13}. The average and median $\log L_{\textrm{rad}}$ [W {Hz}$^{-1}$ sr$^{-1}$] for the sources in this paper were 25.65 and 25.44, respectively (at 0.09$ < z <$ 0.82) while the average $\log L_{\textrm{rad}}$ for sources in \citet{Ka17} was 24.7 W Hz$^{-1}$ sr$^{-1}$ (at redshifts of 0.14$ < z <$ 1.0) and the average $\log L_{\textrm{rad}}$ for sources in \citet{Ga06, Ce13} was 26.45 W Hz$^{-1}$. Sources detected by \citet{Ga06} were more luminous because of the selection of filtering criteria of five sub-areas, which were chosen for detecting only bright extended HyMoRS. The average $\log L_{\textrm{rad}}$ of the sources in this paper was close to the borderline $\log L_{\textrm{rad}}=25.30$ of FR-I and FR-II sources \citep{Fa74}, as expected for HyMoRS. In Fig. \ref{fig:redshift-logL}, we plotted the distribution of radio luminosities of HyMoRS presented in this paper with known redshifts ($z$).
For comparison, we also plotted the radio luminosity of the smallest compact symmetric objects (CSO) \citep{Ku06, Ku10} and low-power compact steep-spectrum sources (CSSs) \citep{Da02a, Da02b} with known redshift. We also included previously detected HyMoRS in \citet{Ka17, Ga06, Ce13}. It was clearly seen in Fig. \ref{fig:redshift-logL} that the sources presented in this paper were less powerful than CSO and CSSs and five HyMoRS by \citet{Ga06, Ce13} but a bit more luminous than previously detected HyMoRS in \citet{Ka17}.

\begin{figure}
\centerline{
\includegraphics[height=11cm,width=8cm,angle=-90]{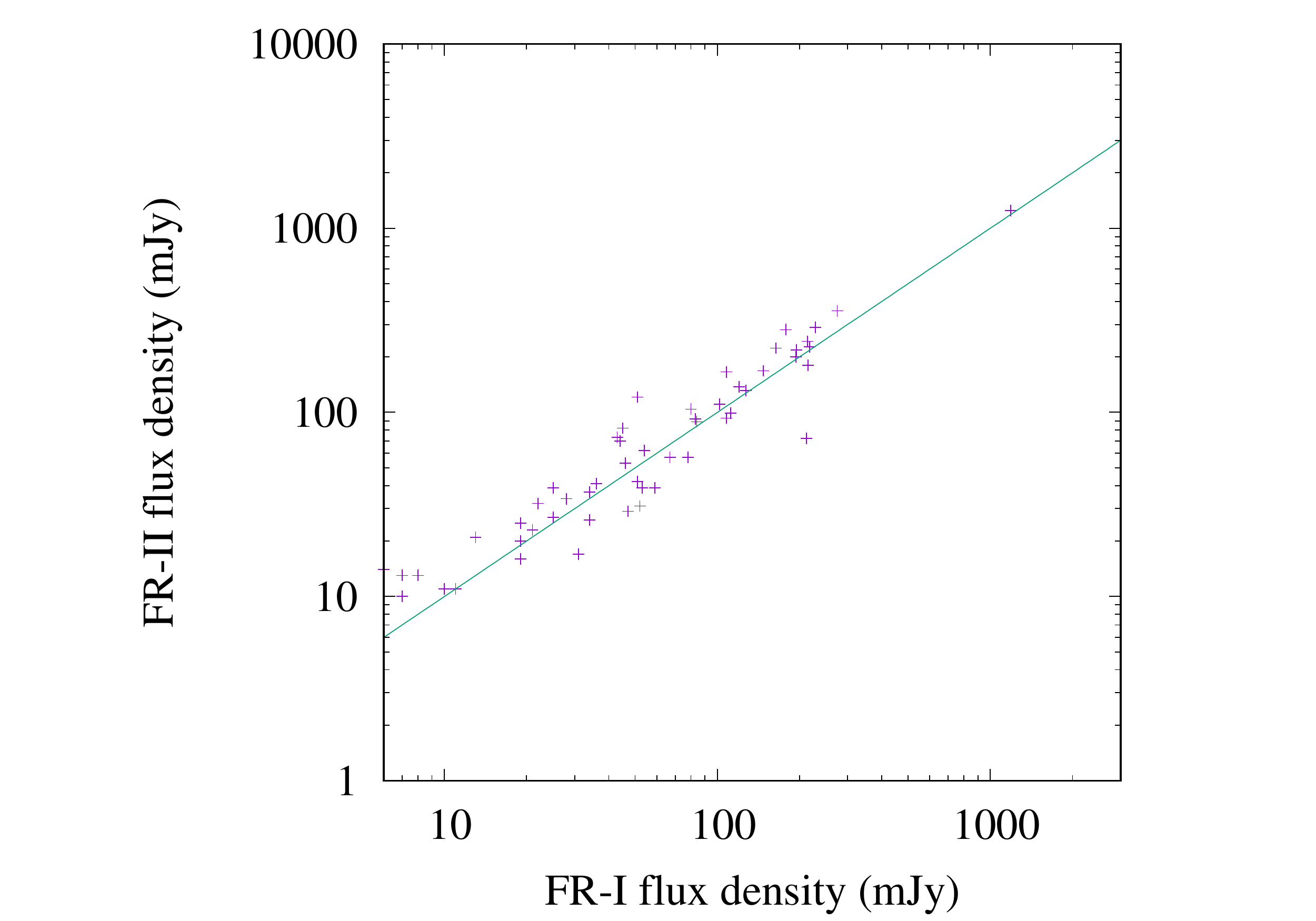}
}
\caption{Flux densities were measured for all HyMoRS of the FR-I and FR-II structures. The error in flux density in FR-I and FR-II was also shown. Here, the green straight line represented the 1:1 flux density between FR-I and FR-II structures.}
\label{fig:fri+frii-flux}
\end{figure}

In Fig. \ref{fig:fri+frii-flux}, we plotted the flux density of FR-I and FR-II lobes for each HyMoRS presented in the current work. Though for most of the galaxies (88 per cent), the FR-II lobe had more flux density compared to the FR-I lobe, as expected for powerful FR-II sources, there were four galaxies (J0715+5528, J1538+2144, J0756+3901, and J2308--0527) (12 per cent) that had FR-I lobe flux density higher than that of the FR-II lobe flux density. For some of them, this might be due to the fact that the flux density of FR-I was boosted due to the relativistic Doppler effect. It was found that the average and median flux density of FR-I were 114 and 53.5 mJy, and that of FR-II were 127 and 62 mJy.

The nature of HyMoRS is yet to be understood. HyMoRS has been used to support the idea that the environment has a significant impact on the formation of radio morphology \citep{Go02, Me08}. The asymmetric structure shown in HyMoRS hints that deceleration may occur on one side of the nucleus of the radio galaxy, causing a different FR morphology.

In our catalogue, four sources (J1236+5525 \citep{Fr17}, J1435+5507 \citep{Sh15}, J1541+4327 \citep{Ch12}, and J1156+3910 \citep{Ch12}) were BRCLGs (Brightest Cluster Galaxies). \citet{Ka17} found one HyMoRS RGZ 122425.8+020310 as a BRCLG as well. This emphasizes the hypothesis that rich environments have an impact on the morphology of HyMoRS \citep{Ka17}. Six sources in our sample (J1229+3137 \citep{Sh15}, J1657+4319 \citep{Ke09}, J1027+1033 \citep{Sh15}, J1336+2329 \citep{Ad08}, J1538+2144 \citep{Sh15}, and J0914+1006 \citep{Ga18}) were LRG (Luminous Red Galaxies), with five sources (J1229+3137, J1657+4319, J1027+1033, J1336+2329, and J1538+2144) having extended structure. The source J0855+4911 \citep{Sh15} was also detected as a bright galaxy. The LRG sources are generally found in high-density regions, which upholds the hypothesis that the jet on either side may be disrupted when it passes through the dense environment \citep{Ce13}. However, the presence of FR-II type hotspots revealed by the higher resolution observations presented by \citet{Ha20} showed that this scenario is not a general one.

In this paper, sources like J1638+3752, J1715+2752, J1657+1945, J1236+5524, J1541+4327, J1249+0932, and J1435+5508 had bent morphology. 
Such bent-jet radio galaxies are more common with FR-I type morphologies [known as wide-angle tails (WATs)] and are thought to be the result of lobes being swept back by the relative motion of the surrounding medium in a dense environment, and they are generally found in rich cluster regions.
WATs are known to have a variety of unusual morphologies, including acceleration regions that are disconnected from the core [e.g. 3C 465; \citep{Od85}]. HyMoRS with bent jets were also found for some of the sources in the recent study \citep{Ha20}. The FR-II like feature in one of the lobes of the bent radio galaxy may be due to the projection effect.
 
The discovery of a large number of HyMoRS in the current work may play a crucial role in understanding the FR-I/FR-II dichotomy, and the discovery of more such objects with future deeper surveys like the LOFAR two-meter sky survey data release II and ASKAP (Australian Square Kilometre Array Pathfinder) is expected. High-resolution multi-wavelength observations of these sources are encouraged to understand the nature of these sources. Future follow-up with radio spectral and polarisation imaging \citep{Ga17} is essential for the confirmation of HyMoRS candidatures of these sources and the study of their inherent asymmetries.

\section{CONCLUSIONS}
\label{sec:conclusion}
At 1400 MHz, we look for hybrid morphology radio sources from the VLA FIRST survey. Our main findings in this paper were as follows:
\begin{description}
\item[$\bullet$] A total of thirty-three HyMoRS were detected, which significantly increases the number of known HyMoRS and opens up the possibility of follow-up deep radio observations of these sources with high-resolution. It may play an important role in understanding the FR dichotomy.
\item[$\bullet$] Except for two sources (which showed flat spectral indices), all HyMoRS showed steep radio spectral indices.		
\item[$\bullet$] Optical counterparts were identified for 29 (16 were from SDSS, 10 were from Pan-STARRS, and 3 were from NED) out of 33 HyMoRS (88 per cent).
\item[$\bullet$] The average value of the FR index for the FR-I and FR-II lobes of HyMoRS in the current sample was comparable to the FR-I and FR-II galaxies in 3CRR sources \citep{La83}.
\item[$\bullet$] The average $\log L_{\textrm{rad}}$ of sources in this paper was close to the borderline $\log L_{\textrm{rad}}$ of FR-I and FR-II sources from \citet{Fa74}, as expected for HyMoRS.
\item[$\bullet$] Out of 33 HyMoRS, there was one quasar, four BRCLG, and six LRG sources. The presence of HyMoRS in high-density regions indicates that the jet on either side may be disrupted when it passes through a dense medium.
\end{description}

\section*{acknowledgments}
This research has made use of the NASA/IPAC Extragalactic Database (NED), which is operated by the Jet Propulsion Laboratory, California Institute of Technology, under contract with the National Aeronautics and Space Administration. This publication makes use of data products from the Two Micron All Sky Survey, which is a joint project of the University of Massachusetts and the Infrared Processing and Analysis Center/California Institute of Technology, funded by the National Aeronautics and Space Administration and the National Science Foundation.

\section*{Data Availability}
We used the data from the catalogue of the VLA FIRST survey \citep{Be95, Wh97}, SDSS Data Release 16 \citep{Ahu20}, PanSTARS \citep{Ch16} and NED. The FIRST Database is publicly available at \href{http://sundog.stsci.edu/index.html}{http://sundog.stsci.edu/index.html}. The data that supports the figures and plots within this paper and the other findings of this study are available from the corresponding author upon reasonable request.


\bsp	
\label{lastpage}
\end{document}